\documentclass[journal,twoside]{IEEEtran}
\usepackage{graphicx}
\usepackage{color,soul}
\usepackage{multirow,booktabs}
\usepackage{cite}

\newcommand{\etal}{et al.}
\usepackage{amsmath}
\usepackage{amsfonts}
\usepackage{amssymb}
\usepackage{cases}
\usepackage{bm}
\usepackage{algorithm}
\usepackage{algorithmic}
\usepackage{balance}
\usepackage{siunitx}
\usepackage{color}
\usepackage{bigstrut}
\usepackage{makecell}
\usepackage{amsthm,amsmath,amssymb}
\usepackage{mathrsfs}

\usepackage{array}

\ifCLASSOPTIONcompsoc
  \usepackage[caption=false,font=normalsize,labelfont=sf,textfont=sf]{subfig}
\else
  \usepackage[caption=false,font=footnotesize]{subfig}
\fi
 \let\MYorigsubfloat\subfloat
 \renewcommand{\subfloat}[2][\relax]{\MYorigsubfloat[]{#2}}
\usepackage{url}

\newcommand{\mathopr}[1]{\mathtt{#1}}

\usepackage{tikz,xcolor,hyperref}
\hypersetup{
    colorlinks=true,
    linkcolor=blue,
    filecolor=blue,
    urlcolor=blue,
    citecolor=blue
}

\definecolor{lime}{HTML}{A6CE39}
\DeclareRobustCommand{\orcidicon}{%
	\begin{tikzpicture}
	\draw[lime, fill=lime] (0,0) 
	circle [radius=0.16] 
	node[white] {{\fontfamily{qag}\selectfont \tiny ID}};
	\draw[white, fill=white] (-0.0625,0.095) 
	circle [radius=0.007];
	\end{tikzpicture}
	\hspace{-3mm}
}

\foreach \x in {A, ..., Z}{%
	\expandafter\xdef\csname orcid\x\endcsname{\noexpand\href{https://orcid.org/\csname orcidauthor\x\endcsname}{\noexpand\orcidicon}}
}

\begin{document}
%
\title{Latent Diffusion, Implicit Amplification: Efficient Continuous-Scale Super-Resolution for Remote Sensing Images}
%
%

\author{Hanlin~Wu\orcidA{},~\IEEEmembership{Member,~IEEE}, Jiangwei~Mo, Xiaohui~Sun, Jie~Ma\orcidB{},~\IEEEmembership{Member,~IEEE}
    \thanks{
        This work was supported in part by the National Natural Science Foundation of China under Grant 62401064 and Grant 62101052, and in part by the Fundamental Research Funds for the Central Universities under Grant 2024TD001. \emph{(Corresponding author: Hanlin Wu.)}

        The authors are with the School of Information Science and Technology, Beijing Foreign Studies University, Beijing 100875, China.
        (e-mail: hlwu@bfsu.edu.cn).}
}

%
%

\markboth{}%
{}
%



\maketitle

\begin{abstract}
    Recent advancements in diffusion models have significantly improved performance in super-resolution (SR) tasks. However, previous research often overlooks the fundamental differences between SR and general image generation. General image generation involves creating images from scratch, while SR focuses specifically on enhancing existing low-resolution (LR) images by adding typically missing high-frequency details. This oversight not only increases the training difficulty but also limits their inference efficiency. Furthermore, previous diffusion-based SR methods are typically trained and inferred at fixed integer scale factors, lacking flexibility to meet the needs of up-sampling with non-integer scale factors. To address these issues, this paper proposes an efficient and elastic diffusion-based SR model (E$^2$DiffSR), specially designed for continuous-scale SR in remote sensing imagery. E$^2$DiffSR employs a two-stage latent diffusion paradigm. During the first stage, an autoencoder is trained to capture the differential priors between high-resolution (HR) and LR images. The encoder intentionally ignores the existing LR content to alleviate the encoding burden, while the decoder introduces an SR branch equipped with a continuous scale upsampling module to accomplish the reconstruction under the guidance of the differential prior. In the second stage, a conditional diffusion model is learned within the latent space to predict the true differential prior encoding. Experimental results demonstrate that E$^2$DiffSR achieves superior objective metrics and visual quality compared to the state-of-the-art SR methods. Additionally, it reduces the inference time of diffusion-based SR methods to a level comparable to that of non-diffusion methods. The code is available at
    \url{https://github.com/hanlinwu/E2DiffSR}.
\end{abstract}


\ifCLASSOPTIONpeerreview
    \begin{center} \bfseries EDICS Category: 3-BBND \end{center}
\fi
%
\IEEEpeerreviewmaketitle

\section{Introduction}

\IEEEPARstart{T}{he} demand for high-resolution (HR) remote sensing images (RSIs) is ever-growing due to their critical role in applications such as land-cover segmentation\cite{xu2023rssformer}, disaster monitoring\cite{zhao2024see}, and urban planning\cite{mao2023elevation}. However, due to the physical constraints of imaging sensors and atmospheric conditions\cite{tu2024rgtgan,wang2024robust}, these images are often captured at low resolutions (LR), which limits their utility. Super-resolution (SR) techniques address this issue by reconstructing HR images from LR inputs, thereby enhancing the details of the data and improving the accuracy of downstream tasks\cite{zhang2023superyolo,liu2023distilling}.

Recent advances in deep learning, including the development of convolutional neural networks (CNNs) and Transformers, have significantly enhanced SR performance in remote sensing\cite{hou2024cswt,xiao2024ttst}. However, two major challenges persist. First, most existing SR models are trained and inferred at fixed integer scale factors (e.g., $\times 2$, $\times 4$, $\times 8$), which limits their flexibility in performing SR tasks with varying scale factors, especially for non-integer scale factors. Training a separate model for each scale factor results in an inefficient use of computational resources and model storage. Second, the visual perceptual quality of SR outputs is often unsatisfactory. Although models optimized with pixel-level losses ($L_1$ or $L_2$) can achieve high peak signal-to-noise ratio (PSNR), they frequently produce blurry and over-smoothed results that lack fine texture details, resulting in poor perceptual quality.

\begin{figure}
    \centering
    \includegraphics[width=\linewidth]{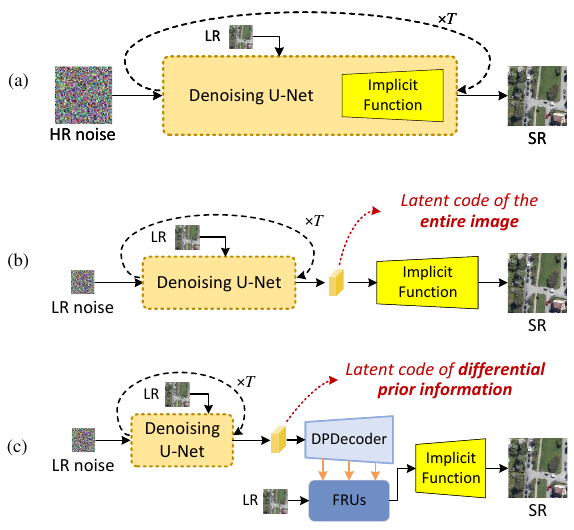}
    \vspace{0em}
    \caption{Model structure comparison with (a) IDM\cite{gao2023implicit}, (b) LDM-based model \cite{kim2024arbitrary}, and (c) our proposed E$^2$DiffSR.}
    \label{fig:intro}
\end{figure}

Continuous-scale SR aims to use a single model to perform SR tasks with arbitrary (integer or non-integer) scale factors. Recent studies have explored various approaches, including architectures based on meta-learning \cite{hu2019meta} and implicit neural representations \cite{chen2021learning,xu2021ultrasr,wu2023learning,chen2023continuous}. Although they can handle different scale factors, the majority of them are optimized by pixel-level losses, leading to lower perceptual quality compared to the state-of-the-art (SOTA) fixed-scale methods.

Generative adversarial networks (GANs)\cite{goodfellow2014generative} have previously dominated the field of perceptual-oriented SR, offering sharper and more visually appealing results\cite{meng2023single,wang2023msagan}. However, GANs are plagued by challenges such as training instability and mode collapse, which can lead to artifacts or inconsistent details. Recently, the introduction of diffusion models (DMs) has provided a new paradigm for various image synthesis tasks, achieving remarkable success in SR\cite{saharia2022image,wang2024sinsr}. However, these models typically perform SR at fixed integer scale factors, which limits their flexibility. Furthermore, DMs are less efficient in SR tasks compared to the GAN-based models, especially with large scale factors, because they operate directly in the high-dimensional pixel space and require a computationally intensive iterative denoising process. Latent diffusion models (LDMs)\cite{rombach2022high} perform the diffusion process in a compressed latent space, offering faster image generation compared to the original DMs while maintaining high-quality output. Therefore, leveraging LDMs to achieve efficient and flexible continuous-scale SR with high perceptual quality is crucial for advancing SR capabilities and expanding their applications in real-world remote sensing tasks.

The most relevant works to ours are implicit diffusion model (IDM) \cite{gao2023implicit} and Yu \etal’s \cite{kim2024arbitrary} LDM-based arbitrary-scale image generation model. IDM combines implicit neural representation with a denoising diffusion model to achieve continuous-scale SR with high perceptual quality. However, IDM operates directly in pixel space, and the larger the image resolution, the more memory and inference time it requires. Later, Yu \etal \cite{kim2024arbitrary} integrates an LDM and an implicit neural decoder to enable efficient and high-quality image generation at arbitrary scales. Although \cite{kim2024arbitrary} improved efficiency, it adopted a general image generation paradigm and ignores the conditional image generation characteristics of the SR task. The general image generation paradigm is inefficient for SR since it requires generating all image details from scratch, leading to unnecessary computational overhead and potential artifacts like pseudo-textures and spatial distortions. Additionally, LDMs strongly rely on a heavyweight autoencoder to achieve compression and restoration from the pixel space to the latent space. However, in scenarios with constrained data, the autoencoder may struggle to learn robust representations, which can severely limit the final performance of the image generation model. This dependency underscores the importance of optimizing the autoencoder's architecture and training process for the SR task.

To address the aforementioned challenges, this article presents E$^2$DiffSR, a novel LDM-based framework specifically designed to provide efficient and elastic continuous-scale SR for RSIs.The proposed E$^2$DiffSR leverages the efficiency of traditional SR methods while harnessing the generative power of diffusion models. By focusing on reconstructing high-frequency details and utilizing existing low-frequency information in the LR image, the proposed LDM achieves a balanced approach that enhances both visual quality and computational efficiency. Additionally, it supports continuous-scale SR, providing flexibility for various remote sensing applications. The main differences between our E$^2$DiffSR and previous works \cite{gao2023implicit,kim2024arbitrary} are shown in Fig.\,\ref{fig:intro}.

Specifically, E$^2$DiffSR adopts a two-stage training paradigm. The first stage is latent space pretraining. Unlike previous methods \cite{rombach2022high,kim2024arbitrary} that compress the entire image into the latent space, we encode only the difference information between the HR and LR images, obtaining a differential prior representation (DPR). Symmetrically, we use a differential prior decoder (DPDecoder) to restore the DPR back to the pixel space. Additionally, we introduce a traditional SR architecture that enriches the representation of the LR image in the LR space through multiple cascaded feature refinement units (FRUs). More importantly, we insert a feature modulation layer before each FRU, allowing the utilization of multi-level outputs from the DPDecoder to assist in reconstructing the missing high-frequency details. Finally, a continuous-scale upsampling module (CSUM) based on implicit neural representation is used to generate the SR output at arbitrary scales. The second stage is the diffusion model learning phase. In this stage, we construct a diffusion model in the latent space, conditioning only on the LR image to generate the DPR. The forward process gradually adds Gaussian noise to the DPR, transitioning it into the standard Gaussian noise distribution space. The reverse process starts from the standard Gaussian noise and iteratively denoises it to generate the predicted DPR. This denoising is implemented by a UNet, which is conditioned on the LR image. At each iteration, the model refines the DPR by progressively removing noise, effectively reconstructing the high-frequency details that were lost in the LR image. Finally, the predicted DPR is combined with the SR model of the first stage, resulting in a super-resolved image that preserves both fine details and overall image coherence.


To the best of our knowledge, E$^2$DiffSR is the first LDM-based continuous-scale SR model for RSIs. The main contributions of this paper are as follows:

\begin{enumerate}
    \item We propose a novel LDM-based continuous-scale SR framework for RSIs, named E$^2$DiffSR. Our proposal reduces the inference time of diffusion-based SR models to be comparable to non-diffusion methods for the first time, while achieving objective metrics and visual quality that surpass the current SOTAs.

    \item We devise a differential autoencoder to compress the difference information between HR and LR images into a compact latent code, tailored to the characteristics of the SR task. Furthermore, we introduce a conditional diffusion model to learn the distribution of the latent space, which greatly improves the learning and inference efficiency of the traditional diffusion model.

    \item We combine the decoder with a traditional SR architecture, allowing the decoded difference information to guide the FRU to complete the reconstruction of high-frequency details, and introduce an implicit function-based upsampling layer to achieve continuous-scale SR. Besides, a scale-aware feature modulation mechanism is introduced to better adapt to multi-task learning.
\end{enumerate}

\section{Related Works}
\label{sec:related_works}

In this section, we first review fixed- and continuous-scale SR methods, respectively. Next, we introduce SR techniques specifically for RSIs. Finally, we provide a brief overview of LDM-based image generation models, which are highly relevant to this article.

\subsection{Fixed-Scale SR}
Recently, deep learning techniques have significantly advanced SR algorithms. We classify deep learning-based SR methods into three main categories: regression-based methods, GAN-based methods, and diffusion-based methods.

\subsubsection{Regression-based Methods}
Inspired by the pioneering work of SR methods based on convolutional neural networks (CNNs) \cite{dong2015image}, subsequent CNN-based models have significantly enhanced SR capabilities. These improvements include the incorporation of residual learning \cite{kim2016accurate, zhang2017beyond}, dense connections \cite{lim2017enhanced, zhang2018residual}, and attention mechanisms \cite{dai2019second,zhu2021lightweight}. Moreover, the advent of transformer-based architectures has brought a paradigm shift to the field by effectively managing long-range dependencies and improving feature fusion. Exemplary models, such as textural transformer network for image SR (TTSR) \cite{yang2020learning} and the hybrid attention transformer (HAT) \cite{chen2023activating}, leverage the self-attention mechanisms inherent in transformer models to achieve superior SR performance.

However, the aforementioned methods are all regression-based, focusing on optimizing pixel-level losses to achieve a high PSNR. These approaches tend to produce oversmoothed outputs and fail to accurately capture fine details and textures. To address these limitations, researchers have focused on optimizing both objective functions and model architectures, leading to the development of GAN- and diffusion-based models that improve visual quality by better capturing realistic textures and details.

\subsubsection{GAN-based Methods}
Building upon the pioneering work of Goodfellow \etal \cite{goodfellow2014generative}, Ledig \etal \cite{ledig2017photo} developed the first perceptual-oriented GAN-based SR model. Wang \etal \cite{wang2018esrgan} advanced the previous architecture by refining adversarial and perceptual loss functions and introducing the residual-in-residual dense block (RRDB) to generate more realistic textures with fewer artifacts. Additionally, the structure-preserving SR (SPSR) introduced by Ma \etal \cite{ma2021structure} addresses structural distortions in GAN-based approaches by incorporating gradient guidance and neural structure extraction, leading to improved recovery of geometric structures and perceptually pleasant details.

Despite their success, GAN-based methods present complex optimization challenges, including the risk of training instability and mode collapse. To address these issues, researchers introduced normalizing flow and diffusion models into SR tasks, providing a more stable and efficient training process while maintaining high image quality.

\subsubsection{Diffusion-based Methods}

Diffusion models have recently emerged as a powerful tool in the field of SR. Building on the foundational work of Ho \etal \cite{ho2020denoising} and Sohl \etal \cite{sohl2015deep}, diffusion models utilize a denoising process to map samples through hidden spaces, allowing for iterative refinement of images. In the context of SR, these models have shown great promise due to their ability to recover fine details and improve image quality beyond GAN-based methods. Notably, SR3 \cite{saharia2022image} pioneered the application of diffusion models for natural scene image SR. Li \etal \cite{li2022srdiff} expanded on this approach by introducing residual prediction during the denoising process. Luo \etal \cite{luo2023image} adopted a continuous-time diffusion model based on stochastic differential equations to complete various image restoration tasks including SR.

Despite the strong visual performance of diffusion-based SR methods, they are primarily designed for fixed-scale SR tasks. While these fixed-factor SR techniques work well at specific integer scaling factors, they lack the flexibility to handle arbitrary upsampling requirements.

\subsection{Continuous-Scale SR}
Continuous scale SR aims to learn a single model to perform SR tasks with arbitrary scale factors to save computational and storage resources.

Hu \etal \cite{hu2019meta} proposed a meta-upsampling module to build a SR network capable of handling arbitrary scales (Meta-SR). Meta-SR creates a coordinate mapping from the LR space to the HR space and uses a fully connected network to dynamically predict the upsampling filter weights for each scale. Behjati \etal \cite{behjati2021overnet} proposed an overscaling module (OSM) that generates overscaled feature maps for continuous-scale upsampling. Later, implicit neural representations that map arbitrary coordinates to pixel values became the mainstream approach for continuous-scale SR. Chen et al. \cite{chen2021learning} introduced local implicit image function (LIIF) for arbitrary-scale SR, which employs a local ensemble strategy to avoid the checkerboard artifact issue. Additionally, UltraSR \cite{xu2021ultrasr} incorporates spatial encoding as a key into the implicit image function, further improving the model performance. Scale-aware dynamic network (SADN) \cite{wu2023learning} proposed a global implicit function for continuous-scale SR, which approximates the continuous representation of an image using a series of discrete feature maps with asymptotic resolutions.

Implicit diffusion model (IDM) \cite{gao2023implicit} is the first method to combine implicit neural representations with diffusion models, extending the capability of arbitrary-scale SR while maintaining high visual quality. However, performing SR with large scale factors using IDM results in extremely low inference efficiency, as it requires repeated denoising in the reverse diffusion process and incurs significant computational overhead by operating directly in the pixel space. Yu \etal \cite{kim2024arbitrary} proposed a continuous-scale SR method based on LDM, which significantly improves inference efficiency by performing diffusion in a lower-dimensional latent space. However, it heavily relies on a large-parameter autoencoder, making efficient training challenging. To address the above issues, we propose a novel architecture to compress the pixel space, allowing the diffusion model to operate in a compact differential prior space. We carefully design a decoder, combining traditional SR architecture with implicit neural representations, to achieve continuous-scale SR.


\subsection{SR Methods for RSIs}

In the field of remote sensing, early SR methods primarily focused on developing innovative neural network architectures to effectively capture the intricate characteristics of RSIs. Lanaras \etal \cite{lanaras2018super} leveraged a deep CNN to enhance Sentinel-2 satellite imagery by super-resolving LR spectral bands. Pan \etal \cite{pan2019super} introduced the backprojection network to enhance RSI SR using global and local residual learning for medium to large scaling factors. Recently, the Transformer architecture have shown remarkable success in capturing long-range dependencies and global context information, have also been adpapted for SR tasks. Xiao \etal \cite{xiao2024ttst} introduced the top-k token selective Transformer (TTST) to tackle challenges in using Transformer-based methods for large-area earth observation in image SR. TTST adaptively eliminates the interference of irrelevant tokens to achieve more compact self-attention calculation, significantly reducing the computational cost and the number of model parameters. However, these PSNR-oriented methods tend to penalize the reconstruction of high-frequency details, thus failing to accurately reflect human visual preferences for the reconstruction results.

To improve the visual perception quality of SR results, many GAN-based methods have been proposed. Lei \etal \cite{lei2019coupled} introduced a GAN-based SR algorithm with a dual-path network, addressing ``discrimination-ambiguity'' in traditional GANs. Dong \etal \cite{dong2020remote} developed a dense-sampling network for large-scale SR in RSIs, using multilevel priors, attention mechanisms, and a chain training strategy to enhance performance. Tu \etal \cite{tu2024rgtgan} introduced a gradient-assisted texture module and dense inner deformable convolution to enhance reference-based SR.

Recently, some studies introduced diffusion models into the SR task of RSI in order to solve the training difficulties of GANs. Wu \etal \cite{wu2023conditional} considered the complex degradations in the real world and leveraged contrastive learning to extract degradation representations, thereby proposing a diffusion-based blind SR method. Xiao \etal \cite{xiao2023ediffsr} proposed an efficient diffusion model for SR of RSIs (EDiffSR), which leveraged an efficient activation network for noise prediction, greatly reducing the computational complexity. Additionally, EDiffSR incorporated a conditional prior enhancement module to improve detail retention, addressing issues of oversmoothing and artifacts.

However, the iterative denoising process of the diffusion model limits the SR efficiency of these approaches, making it essential to transition the diffusion process to a more compact low-dimensional space.

\subsection{LDM-based Image Generation}
LDMs \cite{rombach2022high} have emerged as a groundbreaking approach in the field of generative modeling, particularly for high-resolution image synthesis. By leveraging a two-step process that first compresses high-dimensional data into a lower-dimensional latent space, LDMs significantly enhance computational efficiency while preserving essential details and characteristics of the original images. Subsequently, a diffusion model is applied in this latent space to iteratively denoise and refine the image, ultimately resulting in high-quality outputs.

Ramesh \etal \cite{ramesh2022hierarchical} utilized LDMs for high-resolution text-to-image synthesis, showcasing the model's capability to generate detailed and consistent images from textual descriptions. Kwon \etal \cite{kwon2022diffusion} revealed that pretrained diffusion models inherently possess a well-structured semantic latent space, which can be leveraged for semantic image manipulation. Blattmann \etal \cite{blattmann2023align} extended LDMs for video generation by operating on sequences of latent representations, enabling efficient high-quality video synthesis. In the realm of neuroscience, Takagi \etal \cite{takagi2023high} have utilized LDMs to reconstruct HR images from human brain activity via functional magnetic resonance imaging (fMRI), enhancing our understanding of brain representation and the relationship between computer vision models and human visual systems.

Despite the significant potential of LDMs, a major challenge lies in the initial stage involving the autoencoder, which requires a large model and extensive data for effective training, thereby and directly impacting the performance of the subsequent latent diffusion process. Recent work by Kim \etal \cite{kim2024arbitrary} has advanced arbitrary-scale image generation and upsampling with LDMs, yet it overlooks the critical distinction between SR and general image generation, as SR aims to enhance existing LR images by adding high-frequency details. Recognizing this difference is essential for optimizing LDM architecture specifically for SR tasks to meet their unique demands.

To make LDMs more suitable for SR tasks, our approach simplifies the autoencoder design in the first stage by focusing on encoding the differences between HR and LR images, rather than the entire image content. By capturing only essential differential information, our encoding strategy reduces the complexity of the model and the training difficulty, thus better adapting LDMs to the specific needs of SR tasks.

\begin{figure*}[t]
    \centering
    \includegraphics[width=\textwidth]{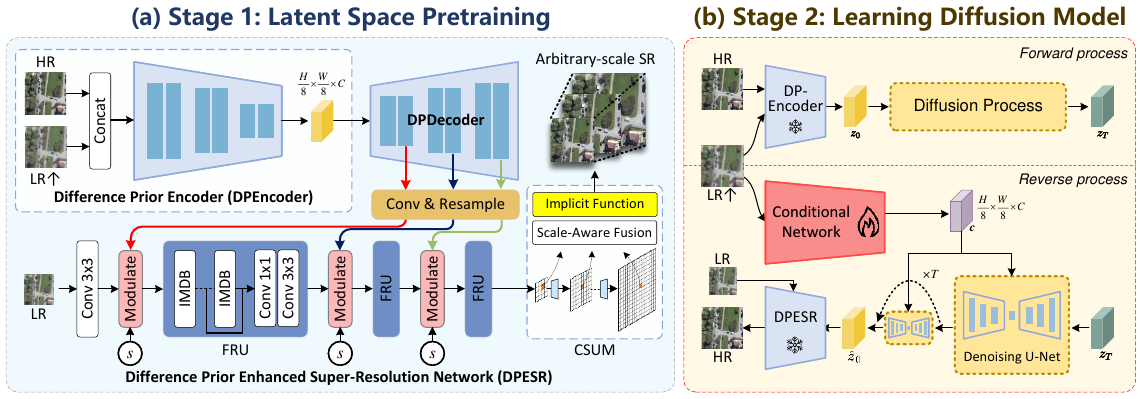}
    \caption{Overview of the proposed E$^2$DiffSR, which consists of two stages: (a) latent space pretraining and (b) learning diffusion model.}
    \label{fig:flowchart}
\end{figure*}

\section{Proposed Method}
\label{sec:methodology}

Although diffusion-based SR methods have shown remarkable advancements, the majority of these models perform the SR task within an image generation paradigm, ignoring the characteristics of the SR task and leading to a waste of computational resources. The core distinction between the SR task and general image generation lies in the fact that SR does not require generating all image information from scratch. Most low-frequency information, such as the type and contours of ground objects, is already present in the LR image. Consequently, SR primarily concentrates on reconstructing the missing high-frequency texture details. Generating all the information from scratch in the SR task can lead to noticeable pseudo-textures and spatial distortions. Inspired by this observation, we developed a novel LDM specifically tailored for SR tasks. This model merges the rapid processing capabilities of traditional SR models with the strong generative capacity of diffusion models, effectively balancing visual quality and computational efficiency. In addition, we introduce a continuous-scale upsampling module (CSUM) so that the proposed model can performe SR with arbitrary integer and non-integer scale factors.

In this section, we first give an overview of our proposed framework in Sec.\,\ref{sec:overview}. Then, the encoder-decoder structure and diffusion model are detailed in Sec.\,\ref{sec:stage1} and Sec.\,\ref{sec:stage2}, respectively. Finally, we summarize the training and inference process of the proposed model in Sec.\,\ref{sec:training_inference}.

\subsection{Overview}
\label{sec:overview}

We propose a novel framework that combines LDMs with classical SR architectures and introduces implicit functions for continuous-scale upsampling. Since the autoencoder in LDM can greatly reduce the computational complexity of the diffusion model, our framework adopts a two-stage training paradigm, the first stage is latent space pretraining, and the second stage uses the diffusion model to learn the probability distribution of the latent space. The flowchart of this framework is shown in Fig.\,\ref{fig:flowchart}.

Considering that LR images already contain a lot of low-frequency information, we propose to encode only the differences between HR and LR images, i.e., high-frequency details, in the first stage. In this way, the autoencoder can disregard the existing content in the LR image, thereby reducing the burden of the pretraining stage. Specifically, we use a differential prior encoder (DPEncoder) that accepts both HR and LR images as input and encodes a compact differential prior.

Then, we construct a differential prior enhanced SR network (DEPSR) to obtain continuous-scale SR results. Specifically, DEPSR consists of three main components: 1) the differential prior decoder (DPDecoder) is used to restore the differential prior to spaces of various resolutions; 2) the SR branch fully utilizes the low-frequency information of the LR image and uses the differential prior as an aid to obtain a feature representation containing rich high-frequency details; 3) the CSUM is employed to achieve arbitrary scale magnification.

It is worth noting that in the inference phase, we do not have HR images to obtain an ideal differential prior. Therefore, we introduce a diffusion model to learn the distribution of differential priors conditioned on LR images, thereby giving an estimate of the differential prior during inference. Since the differential prior is compact and compressed, the denoising process of the diffusion model can be performed in a low-dimensional space. Consequently, the inference speed of the model can be significantly improved compared to previous SR methods based on diffusion models.

\subsection{Encoder-Decoder}
\label{sec:stage1}

The pretraining stage adopts an encoder-decoder architecture, where DPEncoder can be regarded as an encoder and DPESR can be regarded as a decoder.

\subsubsection{DPEncoder} Unlike previous works\cite{rombach2022high,kim2024arbitrary} that directly encode the HR image, DPEncoder only encodes the differential information of HR and LR. Therefore, the input to DPEncoder is a pair of an HR image and its corresponding LR image upsampled using bicubic interpolation, and the output is the differential prior $\mathbf{z}$. The operation of DPEncoder can be summarize as:
\begin{equation}
      \label{eq:yKCf}
      \mathbf{z} = \mathcal{E}_{\mathrm{DP}}(I_{\mathbf{HR}}, I_{\mathbf{LR}}^{\uparrow}),
\end{equation}
where $I_{\mathbf{HR}},I_{\mathbf{LR}}^{\uparrow} \in \mathbb{R}^{H\times W\times 3}$ denote the HR image and the upsampled LR image, respectively. Similar to the vanilla LDM's encoder \cite{rombach2022high}, DPEncoder employs iterative residual blocks and downsampling layers, except that the number of input channels of the first layer is doubled. More specifically, we add three downsampling layers so that the width and height of DPR $\mathbf{z}$ are $1/8$ of the input HR image, i.e., $\mathbf{z}\in\mathbb{R}^{H/8\times W/8\times C}$, where $C$ is the number of channels of the DPR.

\subsubsection{DPESR} DPESR plays the role of the decoder in previous LDMs. In addition to containing a decoder symmetric to DPEncoder, DPESR also includes an SR branch and a CSUM specific to the continuous-scale SR task. Due to this additional SR branch, the process of generating HR images can directly access low-frequency information from LR images, thereby avoiding errors introduced by the encoding-decoding process and significantly reducing the learning difficulty of the latent space pretraining stage.

\subsubsection{DPDecoder} DPDecoder, which is symmetric to DPEncoder, is used to decode compact differential priors $\mathbf{z}$ and contains three upsampling layers. Consequently, DPDecoder can output four decoded differential prior feature maps $\{\mathbf{h}^{(i)}\}_{i=1}^4$ at different resolutions, where $\mathbf{h}^{(i)}\in\mathbb{R}^{H/2^{i-1}\times W/2^{i-1}\times C'}.$ The operation of DPEncoder can be expressed as:
\begin{equation}
      [\mathbf{h}^{(1)}, \mathbf{h}^{(2)}, \mathbf{h}^{(3)}, \mathbf{h}^{(4)}] = \mathcal{D}_{\mathrm{DP}}(\mathbf{z}).
\end{equation}

\subsubsection{SR branch} The SR branch is parallel to DPDecoder and operates similarly to traditional SR networks, functioning within the same resolution space as the LR image to progressively refine coarse LR features.

The SR branch uses a $3\times 3$ convolutional layer as a shallow feature extractor, followed by several sequentially connected feature refinement units (FRUs) to progressively enrich the image representation. To utilize the differential prior obtained from DPDecoder to supplement the missing high-frequency details, we add a modulation layer before each FRU, which takes the differential feature map and the output from the previous FRU as inputs. We use an interpolation operation and a convolution layer to adjust the size and number of channels of the differential feature maps, making them consistent with that of the SR branch.

We observe that the high-frequency differential prior needed for SR tasks varies with different scale factors. Therefore, we use the scale factor $s$ to adjust the ratio of the original features from the FRU to the differential prior from DPDecoder. We use an multi-layer perceptron (MLP) to map $s$ into a series of scaling coefficients,
\begin{equation}
      [\alpha_1^{(1)}, \alpha_2^{(1)}, \alpha_1^{(2)}, \alpha_2^{(2)}, \cdots, \alpha_1^{(4)}, \alpha_2^{(4)}]=\mathopr{MLP}(s),
\end{equation}
where $(\alpha_1^{(i)}, \alpha_2^{(i)})$ represents the coefficient for modulating the input of the $i$th FRU. Then, $\alpha_1^{(i)}$ and $\alpha_2^{(i)}$ are $L_2$-normalized as follows:
\begin{equation}
      [\tilde\alpha_1^{(i)}, \tilde\alpha_2^{(i)}] = \left[
            \frac{|\tilde\alpha_1^{(i)}|}{\sqrt{\tilde\alpha_1^{(i)^2} + \tilde\alpha_2^{(i)^2}+\varepsilon}},
            \frac{|\tilde\alpha_2^{(i)}|}{\sqrt{\tilde\alpha_1^{(i)^2} + \tilde\alpha_2^{(i)^2}+\varepsilon}}
            \right],
\end{equation}
where $\varepsilon$ is a small number to prevent numerical overflow. Finally, the modulation layer before each FRU can be calculated as follows:
\begin{equation}
      \mathbf{\mathbf{f}_{\mathrm{in}}^{(i+1)}} = \alpha_1^{(i)}\cdot \mathbf{f}_{\mathrm{out}}^{(i)} + \alpha_2^{(i)}\cdot \mathopr{Conv}_{3\times 3}(\mathopr{Bicubic}(\mathbf{h}^{(i)})),
\end{equation}
where $\mathbf{f}_{\mathrm{in}}^{(i)}$ and $\mathbf{f}_{\mathrm{out}}^{(i)}$ denote the input and output feature maps of the $i$th FRU, $\mathopr{Conv}_{3\times 3}(\cdot)$ is a  $3\times 3$ convolutional layer, and $\mathopr{Bicubic}(\cdot)$ is the bicubic interpolation.

\begin{figure}[t]
      \centering
      \includegraphics[width=0.95\linewidth]{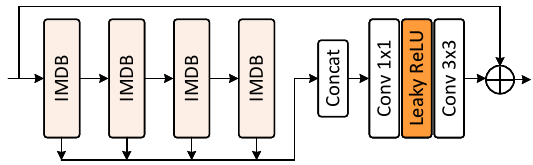}
      \caption{Architecture of the FRU. }
      \label{fig:fru}
\end{figure}

The details of FRU are shown in Fig.\ref{fig:fru}, which consists of four consecutive information multi-distillation blocks (IMDBs)\cite{hui2019lightweight}, and compresses multi-level feature maps through a $1\times 1$ convolution layer. The IMDB is a lightweight feature extraction block specifically designed for SR tasks, effectively utilizing hierarchical features. More details about the IMDB can be found in \cite{hui2019lightweight}.

\subsubsection{CSUM}
To achieve continuous scaling, we introduce a CSUM \cite{wu2023learning} based on implicit functions that can map arbitrary 2-dimensional coordinates to pixel values. CSUM first maps the LR feature map to $T$ feature maps $\{M_t\}_{t=1}^T$ with increasing resolution through multiple pixel-shuffle layers. Then, we employ an implicit function to obtain the continuous representation of the HR image using the information of these $T$ feature maps. The implicit function takes the following form:
\begin{equation}
      s = f_\theta(x; M_1, M_2, \cdots, M_T).
\end{equation}
Specifically, we use a bilinear function to obtain the pixel value $y_i$ of the coordinate $x$ in each feature map $M_i$. Then, we use the scaling factor $s$ to calculate the attention weights to fuse the five interpolation results $\{y_i\}_{i=1}^T$. Finally, the fusion result is fed into a five-layer MLP to obtain the output signal $s$.

\subsection{Diffusion Model for DPR Generation}
\label{sec:stage2}

In the second stage, we learn a diffusion model to simulate the probability distribution of the DPRs. Consequently, in the inference phase, the DPR can be accurately estimated based on the LR image, thereby guiding the DPESR to complete the reconstruction.

\subsubsection{Forward Process}
The forward process transforms the complex data distribution into a Gaussian distribution by gradually adding noise to the data. Specifically, define $\mathbf{z}_0:=\mathcal{E}_{\mathrm{DP}}(I_{\mathrm{HR}}, I_{\mathrm{LR}}^{\uparrow})$ as the ground truth DPR, then
\begin{equation}
      \label{eq:YSqX}
      q(\mathbf{z}_t | \mathbf{z}_{t-1})=\mathcal{N}\Big(\mathbf{z}_t;\sqrt{1-\beta_t}\mathbf{z}_{t-1},\beta_t\mathbf{I}\Big),\quad 1\leq t \leq T,
\end{equation}
where $\{\beta_t\}_{t=1}^T$ is a predefined variance schedule, $T$ is the total time steps of the diffusion process. The noisy latent $\mathbf{z}_t$ at any time step $t$ can be calculated as:
\begin{equation}
      \label{eq:0i6z}
      \mathbf{z}_t =\sqrt{\bar{\alpha}_t}\mathbf{z}_0+\sqrt{1-\bar{\alpha}_t}\bm\epsilon,\quad\bm\epsilon\sim\mathcal{N}(0,\mathrm{I}),
\end{equation}
where $\alpha_t = 1-\beta_t, \bar{\alpha}_t=\prod_{i=0}^t\alpha_i$.

\subsubsection{Reverse Process}
The reverse process starts from a Gaussian noise $\mathbf{z}_T$, iteratively removing the noise to sample an output $\hat{\mathbf{z}}_0$. The transition probability of the reverse process is intractable, but it can be approximated by a neural network as follows:
\begin{equation}
      \label{eq:Boev}
      p_\theta\big(\mathbf{z}_{t-1}|\mathbf{z}_{t}\big)=\mathcal{N}\big(\mathbf{z}_{t-1};\boldsymbol{\mu}_{t}\left(\mathbf{z}_{t},\mathbf{z}_{0}\right),\sigma_{t}^{2}\mathbf{I}\big),
\end{equation}
where $\sigma_t^2=\beta_t$ and
\begin{equation}
      \mu_t(\mathrm{z}_t,\mathrm{z}_0)=\frac{1}{\sqrt{\alpha_t}}\left(\mathrm{z}_t+\frac{\beta_t}{\sqrt{1-\bar{\alpha}_t}}\epsilon\right).
\end{equation}
DDPM \cite{ho2020denoising} proposes to learn a noise predictor $\varepsilon_\theta(\cdot)$ to optimize to reverse process.

\subsubsection{Conditioning Mechanisms}
SR is a typical conditional image generation task, so the denoiser accepts the LR image as a condition. Previous studies \cite{wu2023conditional,xiao2023ediffsr} have shown that extracting features of the LR image as conditions, rather than the original LR image, helps accelerate model convergence. Therefore, we use a conditioning network to extract features of LR images, whose structure is the same as DPEncoder, except that the number of channels in the input layer is reduced to $3$ and only accepts upsampled LR images as input. Besides, for SR tasks with different scale factors, the distribution of DPR between LR and HR images will change significantly. Therefore, we also input the scale factor $s$ as an additional condition into the denoiser. The scale factor $s$ is encoded in the same way as the diffusion time $t$, using the Transformer’s sinusoidal position embedding \cite{vaswani2017attention}. In summary, the estimated noise output by the denoiser at time step $t$ can be calculated as:
\begin{equation}
      \hat{\bm\epsilon} = \bm\epsilon_\theta(\mathbf{z}_t, t, s, \mathopr{CondNet}(I_{\mathrm{LR}}^{\uparrow})),
\end{equation}
where $\mathopr{CondNet}(\cdot)$ is the operation of the conditioning network.

\subsection{Training and Inference}
\label{sec:training_inference}

\subsubsection{Training}
In the pretraining stage, we jointly optimize DPEncoder and DPESR. To ensure that the DPESR generates realistic texture details, we adopt an adversarial learning strategy following \cite{esser2021taming,rombach2022high}, i.e., using a patch-based discriminator to distinguish between the reconstructed image and the original image. Besides, we employ a regularization loss to force the DPEncoder to learn more concise feature representations. We investigated two regularization methods: (i) applying a Kullback-Leibler (KL) divergence constraint as in the variational autoencoder (VAE) \cite{kingma2013auto} to encourage the latent space tend to a standard Gaussian distribution; and (ii) using vector quantization (VQ) to make the distribution of the latent space concentrated on a discrete codebook. In summary, the optimization objective $\mathcal{L}_{\text{S1}}$ of the pretraining stage can be summarized as follows:
\begin{equation}
      \mathcal{L}_{\text{S1}}=\|I_{\mathrm{HR}} - \hat{I}_{\mathrm{HR}}\|_1 + w_1\cdot L_{\mathrm{adv}} + w_2\cdot L_{\mathrm{reg}},
\end{equation}
where $\hat{I}_{\mathrm{HR}} = \mathopr{DPESR}(\mathcal{E}_{\mathrm{DP}}(I_{\mathrm{LR}}, I_{\mathrm{HR}}), I_{\mathrm{LR}})$ represents the reconstructed HR image, $\|\cdot\|_1$ denotes the $L_1$ norm, $L_{\mathrm{adv}}$ and $L_{\mathrm{reg}}$ represent the adversarial loss and regularization loss, respectively. The weights $w_1$ and $w_2$ are the coefficients used to balance the losses.

In the diffusion model training stage, we propose two strategies to make the training more stable: (i) Considering the importance of accurate conditional guidance for SR tasks, we add constraints to the conditioning network to force the output to be close to the ground truth DPR. (ii) We use very few time steps in the reverse process and run all the reverse time steps during the training phase, thereby directly optimizing the $L_1$ loss between the predicted $\hat{\mathbf{z}}_0$ and the ground truth. The training loss of the diffusion modal is defined as follows:
\begin{equation}
      \label{eq:GaHG}
      \mathcal{L}_{\text{S2}} = \|\hat{\mathbf{z}}_0 - \mathbf{z}_0\|_1 + \|\mathopr{CondNet}(I_{\mathrm{LR}}^{\uparrow}) - \mathbf{z}_0\|_1,
\end{equation}
where $\hat{\mathbf{z}}_0$ is obtained iteratively applying Eq.\,\eqref{eq:Boev} and the noise predictor $\bm\epsilon_\theta$. Algorithm \ref{alg:training} summarizes the diffusion model training phase.

\begin{algorithm}[t]
      \renewcommand{\algorithmicrequire}{\textbf{Input:}}
      \renewcommand{\algorithmicensure}{\textbf{Output:}}
      \caption{Diffusion Model Training}
      \label{alg:training}
      \begin{algorithmic}[1]
            \REQUIRE Pretrained DPEncoder and DPESR, total steps $T$, variance schedule $\{\beta_t\}_{t=1}^T$, and training samples $\{I_{\mathrm{HR}}\}$.
            \ENSURE{Trained diffusion model.}
            \STATE{Initialize the parameters of $\bm\epsilon(\cdot)$ and $\mathopr{CondNet}(\cdot)$.}
            \REPEAT
            \STATE{Randomly sample $I_{\mathrm{HR}}$ and a scale foctor $s$ to obtain $(I_{\mathrm{HR}}, I_{\mathrm{LR}})$.}
            \STATE{Compute the DPR: $\mathbf{z} = \mathcal{E}_{\mathrm{DP}}(I_{\mathbf{HR}}, I_{\mathbf{LR}}^{\uparrow})$. (Eq.\eqref{eq:yKCf})}
            \STATE{Randomly sample $\bm\epsilon \sim \mathcal{N}(\bm 0, \bm I)$.}
            \STATE{Compute $\mathbf{z}_T= \sqrt{\bar{\alpha}_T}\mathbf{z}_0+\sqrt{1-\bar{\alpha}_T}\bm\epsilon.$} (Eq. \eqref{eq:0i6z})
            \STATE{Compute the conditional input: $\mathbf{c} = \mathopr{CondNet}(I_{\mathrm{LR}}^{\uparrow})$.}
            \STATE{Let $\hat{\mathbf{z}}_T = \mathbf{z}_T$.}
            \FOR{$t=T, T-1, \cdots, 1$}
            \STATE{$\mathbf{\hat{\mathbf{z}}}_{t-1}=\frac{1}{\sqrt{\alpha_{t}}}\left(\mathbf{\hat{\mathbf{z}}}_{t}-\frac{1-\alpha_{t}}{\sqrt{1-\bar{\alpha}_{t}}}\bm\epsilon_\theta(\hat{\mathbf{z}}_t, t, s, \mathbf{c})\right).$}
            \ENDFOR
            \STATE{Compute loss $\mathcal{L}_{\text{S2}}$ by Eq.\,\eqref{eq:GaHG}.}
            \STATE{Perform a gradient descent step with $\nabla\mathcal{L}_{\text{S2}}$.}
            \UNTIL{converged}
      \end{algorithmic}
\end{algorithm}

\subsubsection{Inference}
In the inference phase, we use the reverse diffusion process to estimate the DPR $\hat{\mathbf{z}}$, and then send it to the DPESR module to obtain SR results with an arbitrary scale factor. Specifically, we first use the conditional network to extract the conditional embedding $\mathbf{c}$ from the LR image. Then, we randomly sample a Gaussian noise $\hat{\mathbf{z}}_T$ and iteratively predict the noisy DPR estimate for the next time step until $\hat{\mathbf{z}}_0$ is obtained. Finally, the estimated DPR and the LR image are sent to the EPESR module to complete the SR reconstruction. Algorithm \ref{algo:inference} summarizes the inference process.

\begin{algorithm}[t]
      \renewcommand{\algorithmicrequire}{\textbf{Input:}}
      \renewcommand{\algorithmicensure}{\textbf{Output:}}
      \caption{Inference}
      \label{algo:inference}
      \begin{algorithmic}[1]
            \REQUIRE{Trained denoiser $\bm\epsilon_\theta$, conditional network $\mathopr{CondNet}$, variance schedule $\{\beta_t\}_{t=1}^T$, LR image $I_{\mathrm{LR}}$, and scale factor $s$.}
            \ENSURE{SR result $I_{\mathrm{SR}}$.}
            \STATE{Compute the LR encoding $\mathbf{c}=\mathopr{CondNet}(I_{\mathrm{LR}}^\uparrow)$.}
            \STATE{Random sample $\hat{\mathbf{z}}_T \sim \mathcal{N}(\bm 0, \bm I)$.}
            \FOR{$t=T, T-1, \cdots, 1$}
            \STATE{Sample $\bm{\xi} \sim \mathcal{N}(\bm 0, \bm I)$}
            \STATE{Predict the noise: $\hat{\bm\epsilon} = \bm\epsilon_\theta(\hat{\mathbf{z}}_t, t, s, \mathbf{c})$.}
            \STATE{
                  Perform an update step by \eqref{eq:Boev},
                  \begin{equation*}
                        \hat{\mathbf{z}}_{t-1}=\frac{1}{\sqrt{\alpha_{t}}}\left(\mathbf{\hat{\mathbf{z}}}_{t}-\frac{1-\alpha_{t}}{\sqrt{1-\bar{\alpha}_{t}}}\hat{\bm\epsilon}\right) + \sqrt{\beta_t}\bm{\xi}.
                  \end{equation*}
            }
            \ENDFOR
            \STATE{Let $\hat{\mathbf{z}} = \hat{\mathbf{z}}_{0}$.}
            \STATE{Compute $I_{\mathrm{SR}}=\mathopr{DPESR}(I_{\mathrm{LR}}, \hat{\mathbf{z}}, s)$.}
      \end{algorithmic}
\end{algorithm}

\section{Experimental Results}
\label{sec:experiments}

In this section, we first introduce the datasets, implementation details, and evaluation metrics. Then, we compare our model with the SOTA continuous-scale and fixed-scale SR methods. Finally, we evaluate the efficiency of the proposed model.

\subsection{Datasets}
We use four public datasets to evaluate the performance of our proposed model, including AID\cite{xia2017aid}, DOTA\cite{xia2018dota}, DIOR\cite{li2020object}, and NWPU-RESISC45\cite{cheng2017remote}. The AID dataset, utilized for both training and testing, consists of 10,000 HR images across 30 aerial scene categories, each with a resolution of $600\times 600$ pixels. We split the AID dataset into training, validation, and test sets in a 9:1:1 ratio, resulting in 9,000 training images, 1,000 validation images, and 1,000 test images. Details on this random split are available in our GitHub repository. For testing, we randomly select 500 images from the DOTA dataset, each with a resolution of $512\times 512$ pixels, and 500 images from the DIOR dataset, each with a resolution of $800\times 800$ pixels. Bicubic interpolation is applied to the AID, DOTA, and DIOR test sets to simulate LR images. Besides, to improve the inference efficiency of the ablation experiments, we randomly selected 30 images from the AID test set and centrally cropped them to a resolution of $512\times 512$ pixels, thereby creating a smaller dataset named AID-tiny.


\subsection{Implementation Details}
In the DPEncoder, we utilize 4 residual groups, each consisting of 2 residual blocks. The base number of channels for the residual blocks is set to 64, and the channel numbers for the four residual groups are set to 1, 2, 2, and 4 times the base channel number, respectively. Each of the first three residual groups is followed by a downsampling layer, reducing the spatial resolution of the DPR to $1/8$ of the input. The number of channels $C$ of the DPR is set to 4. The DPDecoder is symmetrically composed of 4 residual groups, each containing 2 residual blocks. However, unlike the DPEncoder, all residual blocks in these groups have a fixed channel number of 64. The DPESR branch contains 4 FRUs, each composed of 12 IMDBs. The total number of steps $T$ in the diffusion process is set to 4, and the variance schedule linearly decreasing from $\beta_1=0.99$ to $\beta_T=0.1$. The basic channel number in the denoising UNet is set to 64.

During the latent space pretraining stage, we train the autoencoder for 400 epochs with a mini-batch of 4, and each epoch contains 1000 iterations. The learning rate is set to $1.8\times 10^{-5}$. For the initial 5 epochs, only the $L_1$ loss is used for optimization. After this period, a combination of $L_1$ loss, adversarial loss, and regularization loss is employed. The weights $w_1$ and $w_2$ for the adversarial loss and regularization loss are set to $1\times 10^{-6}$ and $0.5$, respectively.

In the second stage, we train the diffusion model for 50 epochs with a mini-batch of 8, and each epoch contains 1000 iterations. The learning rate is set to $8\times 10^{-5}$.

In both training stages, we extract LR patches of size $48\times 48$ as inputs through random cropping. For data augmentation, we randomly flip the images vertically or horizontally and rotate them by 90 degrees. The scale factors for each batch are consistent, and during training, these scale factors are uniformly distributed between 1 and 8. Our E$^2$DiffSR model is implemented using the PyTorch framework and trained on a single 24GB NVIDIA RTX 4090 GPU.

\begin{table*}[htbp]
      \centering
      \renewcommand\arraystretch{1.2}
      \caption{FID\,$\downarrow$ and LPIPS\,$\downarrow$ Scores for Continuous-Scale SR on Three Datasets. The Best Performance is Shown in \textbf{Bold}.}
      \setlength{\tabcolsep}{3.4pt}{
            \begin{tabular}{c|c|cc|cc|cc|cc|cc|cc|cc|cc}
                  \Xhline{4\arrayrulewidth}
                  \multirow{2}[4]{*}{Datasets} & \multirow{2}[4]{*}{Methods} & \multicolumn{2}{c|}{$\times$2.6} & \multicolumn{2}{c|}{$\times$3.0} & \multicolumn{2}{c|}{$\times$3.4} & \multicolumn{2}{c|}{$\times$4.0} & \multicolumn{2}{c|}{$\times$6.0} & \multicolumn{2}{c|}{$\times$8.0} & \multicolumn{2}{c|}{$\times$10.0} & \multicolumn{2}{c}{Average} \bigstrut\\
                  \cline{3-18}          &       & FID   & LPIPS & FID   & LPIPS & FID   & LPIPS & FID   & LPIPS & FID   & LPIPS & FID   & LPIPS & FID   & LPIPS & FID   & LPIPS \bigstrut\\
                  \hline
                  \hline
                  \multirow{5}[2]{*}{AID} & Bicubic & 28.68  & 0.320  & 37.40  & 0.378  & 46.23  & 0.426  & 59.63  & 0.494  & 87.14  & 0.634  & 115.67  & 0.704  & 138.99  & 0.767  & 73.39  & 0.532  \bigstrut[t]\\
                  & LIIF\cite{chen2021learning}  & 9.22  & 0.200  & 16.80  & 0.243  & 27.49  & 0.283  & 42.49  & 0.334  & 80.88  & 0.460  & 112.70  & 0.531  & 136.70  & 0.595  & 60.90  & 0.378  \\
                  & SADN\cite{wu2023learning}  & 8.73  & 0.189  & 15.95  & 0.232  & 25.64  & 0.271  & 39.84  & 0.319  & 76.93  & 0.442  & 109.02  & 0.513  & 131.58  & 0.576  & 58.24  & 0.363  \\
                  & IDM\cite{gao2023implicit}   & 8.70  & 0.160  & 13.83  & 0.208  & 19.97  & 0.233  & 29.25  & 0.277  & 53.92  & 0.369  & 76.95  & 0.425  & \textbf{82.22} & 0.505  & 40.69  & 0.311  \\
                  & Ours  & \textbf{6.92} & \textbf{0.087} & \textbf{9.62} & \textbf{0.113} & \textbf{13.57} & \textbf{0.137} & \textbf{18.37} & \textbf{0.174} & \textbf{33.86} & \textbf{0.257} & \textbf{57.17} & \textbf{0.353} & 82.31  & \textbf{0.465} & \textbf{31.69} & \textbf{0.227} \bigstrut[b]\\
                  \hline
                  \multirow{5}[2]{*}{DOTA} & Bicubic & 43.10  & 0.238  & 56.37  & 0.289  & 69.47  & 0.335  & 82.63  & 0.399  & 114.33  & 0.538  & 147.56  & 0.612  & 170.82  & 0.672  & 97.75  & 0.440  \bigstrut[t]\\
                  & LIIF\cite{chen2021learning}  & 16.53  & 0.133  & 30.12  & 0.170  & 43.22  & 0.203  & 58.78  & 0.246  & 100.55  & 0.355  & 134.14  & 0.423  & 160.71  & 0.486  & 77.72  & 0.288  \\
                  & SADN\cite{wu2023learning}  & 15.70  & 0.128  & 27.87  & 0.163  & 40.33  & 0.194  & 55.12  & 0.235  & 95.92  & 0.339  & 129.09  & 0.402  & 154.11  & 0.463  & 74.02  & 0.275  \\
                  & IDM\cite{gao2023implicit}   & 15.26  & 0.111  & 24.31  & 0.145  & 32.97  & 0.171  & 44.81  & 0.207  & 77.86  & 0.298  & 107.60  & \textbf{0.354} & \textbf{120.01} & \textbf{0.431} & 60.40  & 0.245  \\
                  & Ours  & \textbf{11.55} & \textbf{0.069} & \textbf{18.73} & \textbf{0.091} & \textbf{25.68} & \textbf{0.114} & \textbf{33.75} & \textbf{0.151} & \textbf{61.93} & \textbf{0.249} & \textbf{104.54} & 0.355  & 129.71  & 0.460  & \textbf{55.13} & \textbf{0.213} \bigstrut[b]\\
                  \hline
                  \multirow{5}[2]{*}{DIOR} & Bicubic & 29.71  & 0.315  & 38.97  & 0.375  & 49.75  & 0.426  & 65.41  & 0.499  & 109.10  & 0.650  & 131.50  & 0.720  & 150.50  & 0.786  & 82.13  & 0.539  \bigstrut[t]\\
                  & LIIF\cite{chen2021learning}  & 12.70  & 0.202  & 16.03  & 0.250  & 21.50  & 0.292  & 34.38  & 0.349  & 89.80  & 0.487  & 124.24  & 0.561  & 151.65  & 0.625  & 64.33  & 0.395  \\
                  & SADN\cite{wu2023learning}  & \textbf{12.04} & 0.195  & 15.30  & 0.241  & 20.53  & 0.282  & 32.61  & 0.336  & 85.26  & 0.474  & 119.95  & 0.547  & 148.12  & 0.611  & 61.97  & 0.384  \\
                  & IDM\cite{gao2023implicit}   & 13.69  & 0.158  & 16.23  & 0.205  & 19.28  & 0.236  & 26.98  & 0.284  & 64.73  & 0.392  & 97.36  & 0.452  & 108.48  & 0.535  & 49.54  & 0.323  \\
                  & Ours  & 13.81  & \textbf{0.097} & \textbf{15.53} & \textbf{0.131} & \textbf{17.67} & \textbf{0.161} & \textbf{23.03} & \textbf{0.212} & \textbf{46.77} & \textbf{0.300} & \textbf{67.31} & \textbf{0.387} & \textbf{94.75} & \textbf{0.477} & \textbf{39.84} & \textbf{0.252} \bigstrut[b]\\
                  \Xhline{4\arrayrulewidth}
            \end{tabular}%
      }
      \label{tab:cssr-results}%
\end{table*}%

\begin{figure*}[htbp]
      \begin{minipage}{0.495\textwidth}
            \centering
            \includegraphics[width=\linewidth]{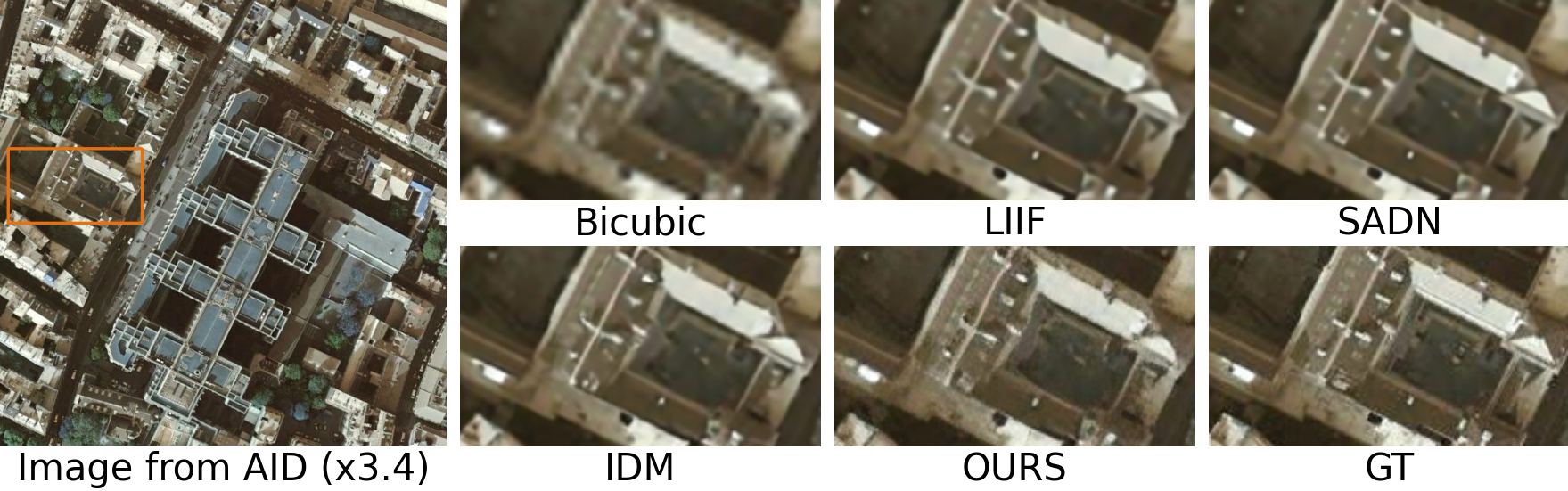} 
      \end{minipage}
      \hfill
      \begin{minipage}{0.495\textwidth}
            \centering
            \includegraphics[width=\linewidth]{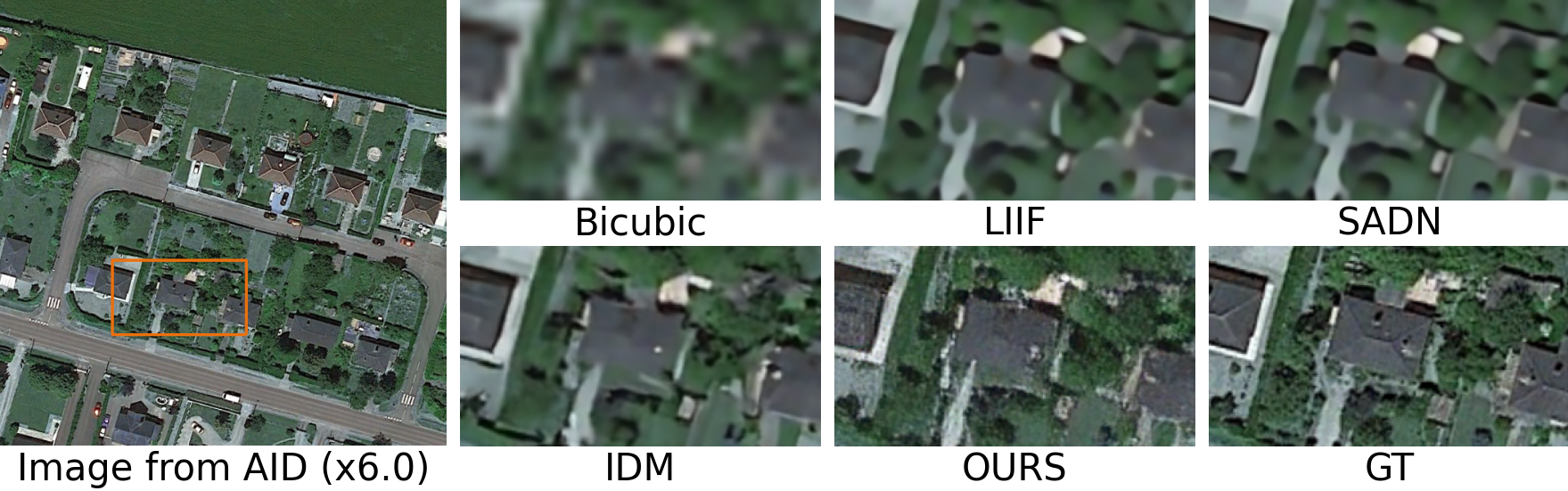} 
      \end{minipage}

      \vspace{5pt} 

      \begin{minipage}{0.495\textwidth}
            \centering
            \includegraphics[width=\linewidth]{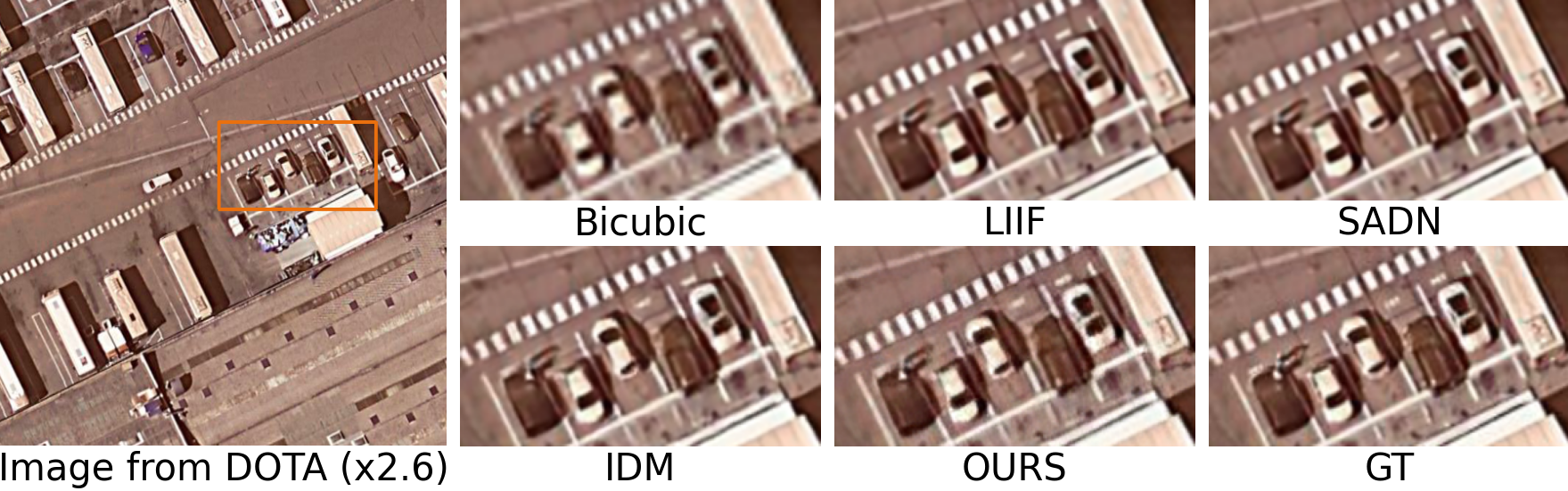} 
      \end{minipage}
      \hfill
      \begin{minipage}{0.495\textwidth}
            \centering
            \includegraphics[width=\linewidth]{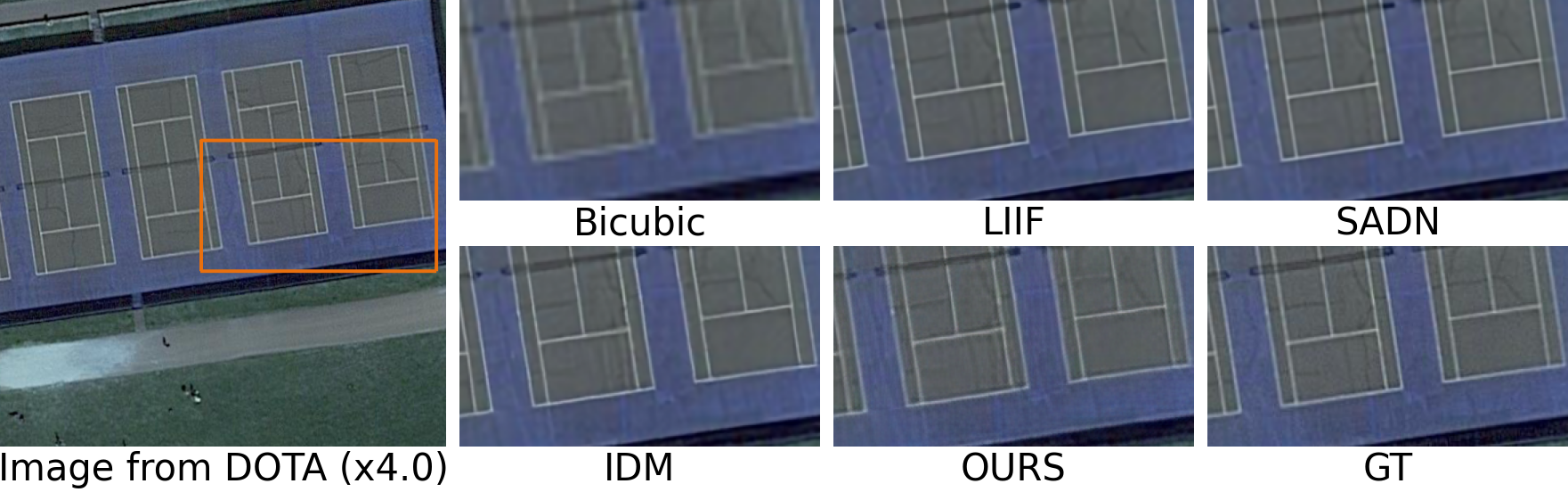} 
      \end{minipage}

      \vspace{5pt} 

      \begin{minipage}{0.495\textwidth}
            \centering
            \includegraphics[width=\linewidth]{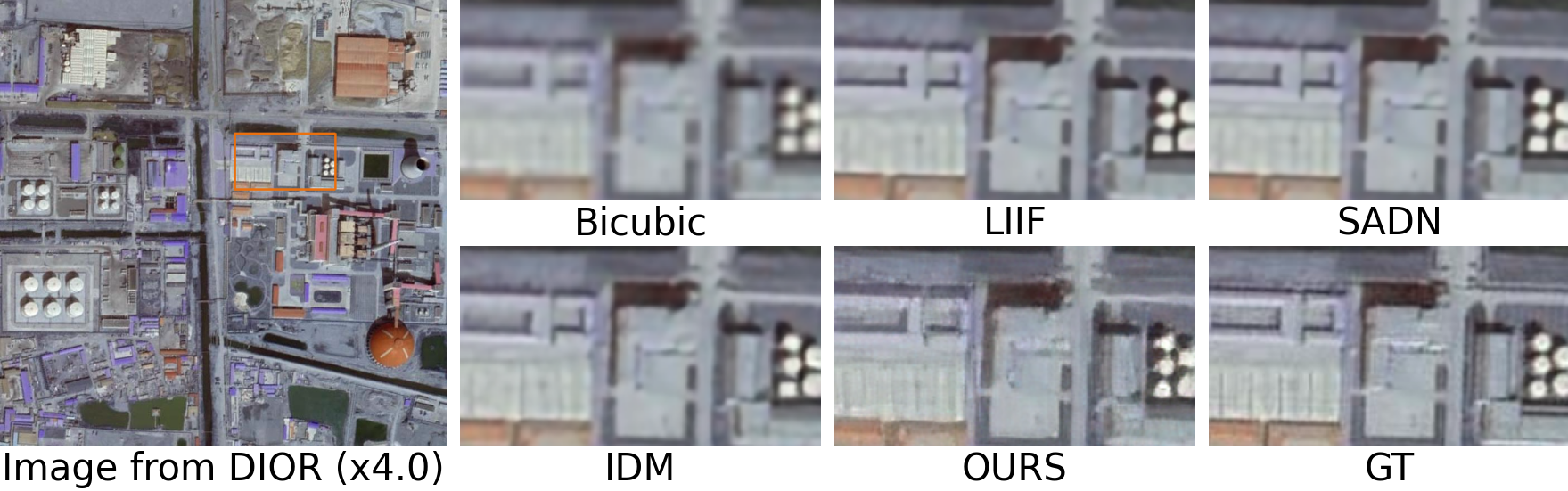} 
      \end{minipage}
      \hfill
      \begin{minipage}{0.495\textwidth}
            \centering
            \includegraphics[width=\linewidth]{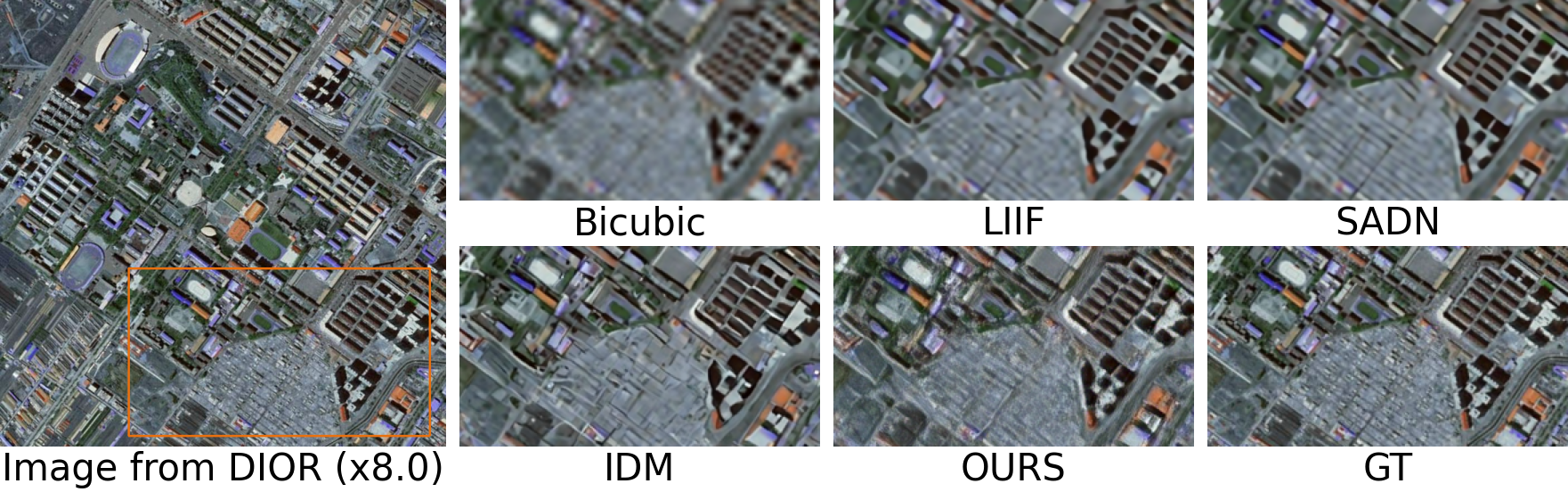} 
      \end{minipage}

      \caption{Visual comparisons between continuous-scale SR methods. Zoomed-in view for a better view.}
      \label{fig:main-continuous-scale}
\end{figure*}

\subsection{Evaluation Metrics}
To comprehensively evaluate the performance of our E$^2$DiffSR model, we utilize four metrics: PSNR, learned perceptual image patch similarity (LPIPS)\cite{zhang2018unreasonable}, and Fréchet inception distance (FID)\cite{heusel2017gans}.

PSNR is commonly used to quantify the pixel-wise accuracy and structural similarity between the super-resolved images and their corresponding ground truths. For perceptual quality assessment, we use LPIPS, which evaluates similarity based on features extracted by deep neural networks, thereby aligning more closely with human visual perception. FID is used to measure the statistical similarity between the distribution of super-resolved images and that of real HR images, offering insights into the overall quality and realism of the generated outputs.

\begin{table*}[t]
      \centering
      \caption{Comparison of PSNR(\si{\dB}) $\uparrow$, LIPIPS $\downarrow$ and FID $\downarrow$ for Integer Scale Factors ($\times$2, $\times$4, $\times$8) on Three Test Datasets. \textbf{Bold} Indicates the Best Results Among all Continuous-Scale Models.}
      \renewcommand\arraystretch{0.95}
      \setlength{\tabcolsep}{7pt}{
            \begin{tabular}{ccc|ccc|ccc|ccc}
                  \Xhline{4\arrayrulewidth}
                  \multirow{2}[4]{*}{Category} & \multirow{2}[4]{*}{Methods} & \multirow{2}[4]{*}{Metric} & \multicolumn{3}{c|}{AID} & \multicolumn{3}{c|}{DOTA} & \multicolumn{3}{c}{DIOR} \bigstrut\\
                  \cline{4-12}          &       &       & $\times$2    & $\times$4    & $\times$8    & $\times$2    & $\times$4    & $\times$8    & $\times$2    & $\times$4    & $\times$8 \bigstrut\\
                  \hline
                  \hline
                  \multirow{3}[2]{*}{Baseline} & \multirow{3}[2]{*}{Bicubic} & PSNR  & 32.498  & 27.430  & 24.502  & 35.983  & 30.247  & 26.783  & 32.484  & 26.948  & 23.982  \bigstrut[t]\\
                  &       & LPIPS & 0.199  & 0.494  & 0.704  & 0.139  & 0.399  & 0.612  & 0.192  & 0.499  & 0.720  \\
                  &       & FID   & 15.726  & 59.633  & 115.667  & 23.179  & 82.629  & 147.557  & 18.712  & 65.415  & 131.497  \bigstrut[b]\\
                  \hline
                  \multicolumn{1}{c}{\multirow{12}[8]{*}{\makecell{Fixed-scale\\ models}}} & \multirow{3}[2]{*}{HAT-L\cite{chen2023activating}} & PSNR  & 35.482  & 29.490  & 25.802  & 39.882  & 33.465  & 28.730  & 35.826  & 29.179  & 25.321  \bigstrut[t]\\
                  &       & LPIPS & 0.115  & 0.321  & 0.515  & 0.073  & 0.236  & 0.407  & 0.118  & 0.337  & 0.548  \\
                  &       & FID   & 4.345  & 39.818  & 110.007  & 4.486  & 55.628  & 130.023  & 7.397  & 32.679  & 120.205  \bigstrut[b]\\
                  \cline{2-12}          & \multirow{3}[2]{*}{SR3\cite{saharia2022image}} & PSNR  & 34.049  & 28.150  & 20.608  & 38.199  & 31.299  & 20.405  & 34.147  & 27.674  & 21.001  \bigstrut[t]\\
                  &       & LPIPS & 0.071  & 0.252  & 0.397  & 0.050  & 0.200  & 0.373  & 0.069  & 0.268  & 0.433  \\
                  &       & FID   & 4.271  & 26.014  & 58.758  & 4.839  & 41.593  & 91.964  & 8.243  & 25.550  & 79.398  \bigstrut[b]\\
                  \cline{2-12}          & \multirow{3}[2]{*}{EDiffSR\cite{xiao2023ediffsr}} & PSNR  & 30.913  & 25.238  & 22.956  & 35.192  & 27.881  & 24.471  & 32.515  & 24.923  & 22.705  \bigstrut[t]\\
                  &       & LPIPS & 0.055  & 0.273  & 0.417  & 0.057  & 0.301  & 0.483  & 0.056  & 0.332  & 0.459  \\
                  &       & FID   & 5.508  & 25.968  & 59.600  & 10.003  & 53.119  & 110.961  & 9.959  & 32.461  & 79.950  \bigstrut[b]\\
                  \cline{2-12}          & \multirow{3}[2]{*}{SPSR\cite{ma2021structure}} & PSNR  & 32.602  & 25.948  & 21.002  & 36.797  & 28.498  & 22.599  & 33.013  & 25.536  & 20.812  \bigstrut[t]\\
                  &       & LPIPS & 0.044  & 0.185  & 0.371  & 0.033  & 0.170  & 0.357  & 0.047  & 0.203  & 0.405  \\
                  &       & FID   & 5.035  & 20.943  & 52.618  & 6.206  & 44.311  & 96.701  & 9.389  & 30.596  & 82.822  \bigstrut[b]\\
                  \hline
                  \hline
                  \multicolumn{1}{c}{\multirow{12}[8]{*}{\makecell{Continious-scale\\ models}}} & \multirow{3}[2]{*}{LIIF\cite{chen2021learning}} & PSNR  & 35.240  & 29.323  & 25.706  & 38.934  & 32.840  & 28.447  & 35.625  & 29.011  & 25.243  \bigstrut[t]\\
                  &       & LPIPS & 0.122  & 0.334  & 0.531  & 0.077  & 0.246  & 0.423  & 0.123  & 0.349  & 0.561  \\
                  &       & FID   & 4.600  & 42.490  & 112.695  & 4.880  & 58.776  & 134.135  & 7.864  & 34.375  & 124.239  \bigstrut[b]\\
                  \cline{2-12}          & \multirow{3}[2]{*}{SADN\cite{wu2023learning}} & PSNR  & \textbf{35.585} & \textbf{29.583} & \textbf{25.892} & \textbf{39.958} & \textbf{33.510} & \textbf{28.855} & \textbf{35.862} & \textbf{29.247} & \textbf{25.411} \bigstrut[t]\\
                  &       & LPIPS & 0.115  & 0.319  & 0.513  & 0.073  & 0.235  & 0.402  & 0.118  & 0.336  & 0.547  \\
                  &       & FID   & \textbf{4.348} & 39.838  & 109.020  & \textbf{4.605} & 55.118  & 129.093  & \textbf{7.355} & 32.611  & 119.949  \bigstrut[b]\\
                  \cline{2-12}          & \multirow{3}[2]{*}{IDM\cite{gao2023implicit}} & PSNR  & 30.730  & 27.611  & 24.216  & 34.925  & 31.494  & 27.062  & 31.542  & 27.557  & 23.790  \bigstrut[t]\\
                  &       & LPIPS & 0.105  & 0.277  & 0.425  & 0.067  & 0.207  & \textbf{0.354} & 0.101  & 0.284  & 0.452  \\
                  &       & FID   & 6.596  & 29.248  & 76.955  & 7.604  & 44.807  & 107.596  & 10.979  & 26.985  & 97.358  \bigstrut[b]\\
                  \cline{2-12}          & \multirow{3}[2]{*}{Ours} & PSNR  & 32.912  & 27.186  & 23.744  & 37.239  & 30.717  & 26.230  & 32.768  & 26.509  & 23.039  \bigstrut[t]\\
                  &       & LPIPS & \textbf{0.053} & \textbf{0.174} & \textbf{0.353} & \textbf{0.039} & \textbf{0.151} & 0.355  & \textbf{0.055} & \textbf{0.212} & \textbf{0.387} \\
                  &       & FID   & 5.735  & \textbf{18.372} & \textbf{57.167} & 6.141  & \textbf{33.753} & \textbf{104.545} & 10.022  & \textbf{23.030} & \textbf{67.309} \bigstrut[b]\\
                  \Xhline{4\arrayrulewidth}
            \end{tabular}%
      }
      \label{tab:fssr-results}%
\end{table*}%

\subsection{Continuous-Scale SR Comparison}
We take bicubic interpolation as the baseline and compare our E$^2$DiffSR with three SOTA continuous-scale SR methods, including LIIF\cite{chen2021learning}, IDM\cite{gao2023implicit} and SADN\cite{wu2023learning}. LIIF\cite{chen2021learning} is a pioneering continuous-scale SR method based on implicit neural representations. SADN addresses the challenges of continuous-scale SR in RSIs by integrating dynamic scale awareness with global implicit functions. LIIF and SADN are both PSNR-oriented models, whereas IDM\cite{gao2023implicit} focuses on perceptual quality and is the first continuous-scale SR method based on diffusion models. To ensure a fair comparison, we retrained these models from scratch on the AID dataset using their official implementations.

\subsubsection{Quantitative Comparison}
Table \ref{tab:cssr-results} presents the comparison of FID and LPIPS metrics of the proposed E$^2$DiffSR and the competing algorithms on three test dataset. We randomly selected 7 scale factors for evaluation. The proposed method consistently outperforms all other approaches across various scale factors, showcasing a clear advantage in both generative quality and perceptual similarity.

We observe that the diffusion-based models achieve overall better performance than traditional PSNR-oriented methods. Specifically, at medium scale factors (e.g., $\times 3.0$ to $\times 8.0$), E$^2$DiffSR outperforms the closest competitor, IDM, by a substantial margin in both FID and LPIPS scores. For the challenging scale factor of $\times 10.0$, where the loss of information from LR images makes SR particularly difficult, our algorithm achieves results that are comparable to or even better than IDM. Besides, for smaller scale factors, such as $\times 2.6$, traditional $L_1$ loss-based algorithms (e.g., SADN) can still yield satisfactory results. This is largely because LR images with smaller scale factors retain most of the necessary reconstruction information, placing lower demands on the model's generative capabilities. Overall, the results convincingly illustrate the superiority of E$^2$DiffSR in achieving high perceptual quality continuous-scale SR results for RSIs.

\subsubsection{Visual Comparison}
We compare the visual results of the continuous-scale SR methods in Fig.\,\ref{fig:main-continuous-scale}. For each test dataset, we randomly selected SR results with two different scale factors for display. Our proposed E$^2$DiffSR excels in generating more realistic SR images compared to traditional methods that optimized by pixel-wise losses, such as LIIF and SADN. These conventional techniques often produce results that are blurry and overly smooth, particularly at larger scaling factors. For instance, in the case of images from the AID dataset with a scaling factor of $\times$6, our method significantly improves the representation of natural elements, such as trees, showcasing intricate details and textures that enhance the overall visual quality. This comparison highlights the advantages of the proposed E$^2$DiffSR in achieving superior SR performance for RSIs, ensuring more accurate and visually appealing results.

\subsection{Fixed-Scale SR Comparison}
We compare our proposed E$^2$DiffSR with four fixed-scale SR methods, including HAT-L \cite{chen2023activating}, SPSR \cite{ma2021structure}, SR3 \cite{saharia2022image}, and EDiffSR \cite{xiao2023ediffsr}; and three continuous-scale SR methods, including LIIF \cite{chen2021learning}, SADN \cite{wu2023learning}, and IDM \cite{gao2023implicit}, on three benchmark datasets. Among these fixed-scale SR models, HAT-L is a SOTA algorithm based on the Transformer architecture, demonstrating advanced capabilities in feature representation. Meanwhile, SR3 is a pioneering approach to image SR using diffusion models, while EDiffSR is a recent approach that utilizes diffusion models for SR in RSIs, delivering high-quality results with enhanced efficiency. In addition, SPSR addresses the perceptual quality issue in SR, but employs a GAN framework in its learning process. The results of these fixed-scale SR models are obtained by retraining on each scale factor.

\begin{figure*}[htbp]
      \centering\scriptsize
      \begin{minipage}[t]{0.04\linewidth}
            \vspace{-150pt}
            $\times$2
      \end{minipage}%
      \hfill
      \begin{minipage}[t]{0.96\linewidth}
            \includegraphics[width=\linewidth]{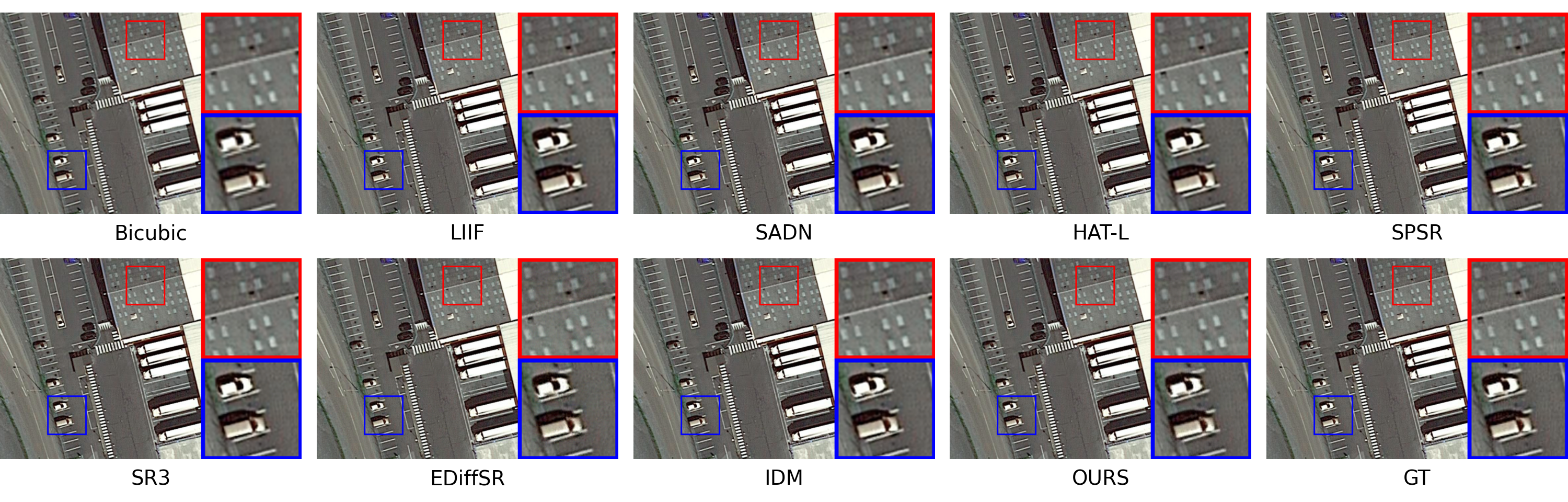}
      \end{minipage}

      \vspace{-6pt}\rule{\textwidth}{0.5pt}\vspace{2pt}

      \begin{minipage}[t]{0.04\linewidth}
            \vspace{-150pt}
            $\times$4
      \end{minipage}%
      \hfill
      \begin{minipage}[t]{0.96\linewidth}
            \includegraphics[width=\linewidth]{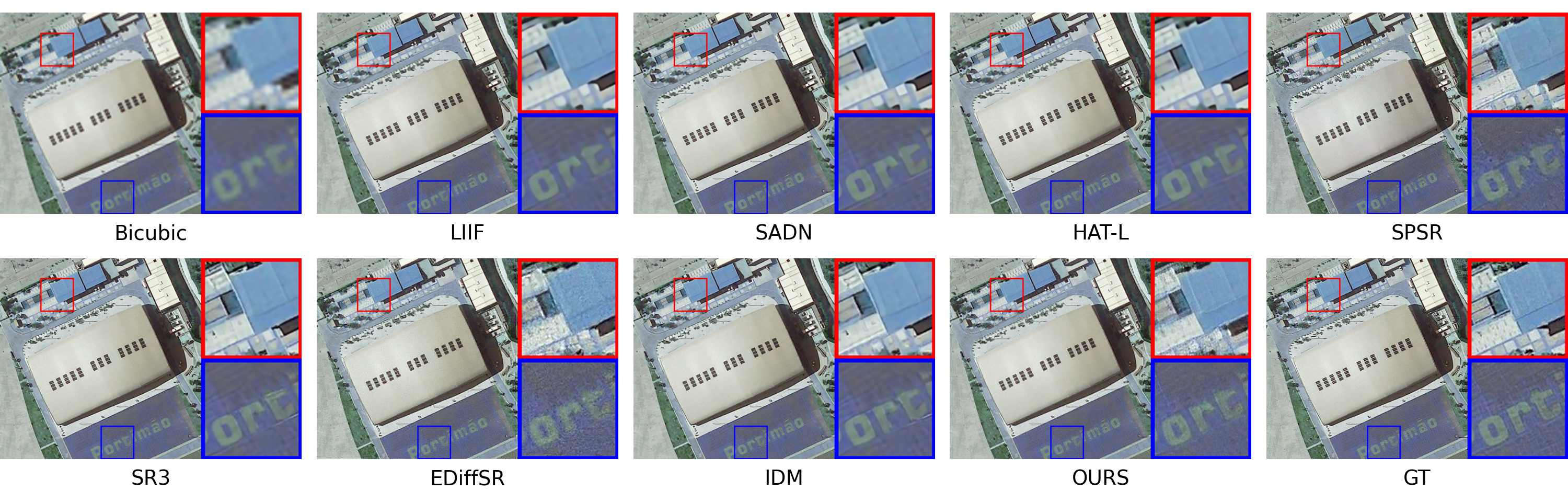}
      \end{minipage}

      \vspace{-6pt}\rule{\textwidth}{0.5pt}\vspace{2pt}

      \begin{minipage}[t]{0.04\linewidth}
            \vspace{-150pt}
            $\times$8
      \end{minipage}%
      \hfill
      \begin{minipage}[t]{0.96\linewidth}
            \includegraphics[width=\linewidth]{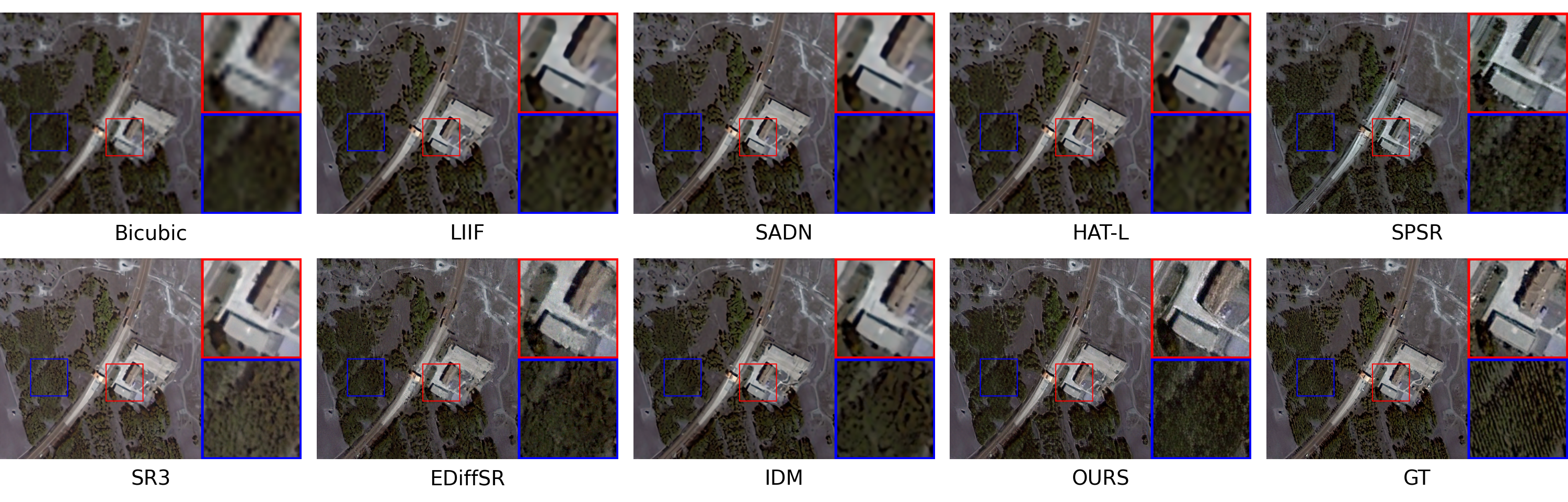}
      \end{minipage}
      \caption{Visual comparison of the fixed-/continuous-scale SR methods with integer scale factors of $\times$2, $\times$4, and $\times$8. Zoomed-in view for a better view. }
      \label{fig:main-fixed-scale}
\end{figure*}

\subsubsection{Quantitative Comparison}
Table\,\ref{tab:fssr-results} presents a comparison of different methods across three test datasets using three integer scale factors ($\times$2, $\times$4, $\times$8), evaluated based on PSNR, LPIPS, and FID metrics. Among all fixed-scale models, HAT-L achieved the highest PSNR; however, it significantly lagged behind GAN- and diffusion-based methods in terms of visual perceptual quality metrics. Among the continuous-scale models, the regression-based method SADN attained the best PSNR. While our proposed method achieved a PSNR comparable to that of IDM, outperforming other GAN- or diffusion-based methods. This underscores our method's ability to effectively balance the trade-off between visual perceptual quality and spatial distortion in SR tasks. Besides, our E$^2$DiffSR demonstrates clear advantages in LPIPS and FID scores, particularly at larger scale factors ($\times$4, $\times$8), indicating its superior ability to learn the data distribution of HR samples and generate high-quality SR results.

\subsubsection{Visual Comparison}
In Fig.\,\ref{fig:main-fixed-scale}, we present a visual comparison of all competing methods at integer scale factors. Our E$^2$DiffSR demonstrates superior SR results, surpassing SOTA methods for both continuous and fixed scale factors.

The results in the first two rows correspond to the scale factor of $\times$2, where the visual differences among the various methods are minimal, and all approaches deliver satisfactory outcomes for this relatively simple SR task. However, a more pronounced distinction emerges in the middle two rows for the scale factor of $\times$2. Here, methods like LIIF, SADN, and \mbox{HAT-L} produce smoother results that, while having emphasized edges, lack essential texture details. Conversely, SPSR and EDiffSR attempt to enhance detail but introduce significant noise and inconsistent artifacts compared to the ground truth.

For the final two rows at the scale factor of $\times$8, our proposed method displays a remarkable ability to provide more natural and realistic results. Not only do the textures in the forested areas appear richer and more authentic, but our method also excels in reconstructing the edges of buildings and other structures effectively. This illustrates that our approach successfully leverages the strengths of both traditional SR models and diffusion models, enhancing detail generation capabilities while maintaining low spatial distortion.





\begin{table}[tbp]
    \centering
    \caption{Comparison of Average Inference Time and Number of Parameters for Various SR Methods. Inference Time is Tested on the AID-tiny Dataset.}
    \label{tab:runtime}
    \renewcommand\arraystretch{1.2}
    \setlength{\tabcolsep}{2.5pt}{
        \begin{tabular}{llcc}
            \Xhline{4\arrayrulewidth}
            Category & Methods & Runtime (s) & \# Param. (M) \bigstrut\\
            \hline
            \multicolumn{1}{l}{\multirow{3}[2]{*}{Regression-based}} & LIIF  & 0.0575 & 22.3 \bigstrut[t]\\
            & SADN  & 0.0581 & 7.6 \\
            & HAT-L &  0.2474 & 40.32 \bigstrut[b]\\
            \hline
            GAN-based & SPSR  & 0.1216 & 24.79 \bigstrut\\
            \hline
            \multicolumn{1}{l}{\multirow{6}[2]{*}{Diffusion-based}} & SR3 (40 steps) & 2.8150 & 92.56 \bigstrut[t]\\
            & SR3-vanilla (1000 steps) &  143.1442  & 92.56 \\
            & EDiffSR & 13.5648 & 26.79 \\
            & IDM (40 steps) & 3.0883 & 111.34 \\
            & IDM-vanilla (1000 steps) & 160.0196  & 111.34 \\
            & Ours  & 0.2082 & 31.25 \bigstrut[b]\\
            \Xhline{4\arrayrulewidth}
        \end{tabular}%
    }
\end{table}%

\subsection{Model Complexity Comparison}

We evaluate the model efficiency by comparing the inference times across different methods on the AID-tiny dataset. Since SR3 and IDM can set different numbers of steps in the reverse diffusion process to speed up inference, we include the inference time of these two models with 40 accelerated steps and full 2,000 steps in the comparison. Table \ref{tab:runtime} presents a comparison of the average inference times and the number of parameters for each method. The results were tested on a server equipped with an NVIDIA GeForce RTX 4090 graphics card, an AMD EPYC 9754 128-core CPU, and 60GB of RAM.

In terms of inference time, regression-based and GAN-based models demonstrate high efficiency, while methods based on diffusion models exhibit significantly longer inference times. This extended duration is primarily due to the iterative denoising associated with reverse diffusion process. Furthermore, the number of reverse denoising steps in diffusion models correlates positively with inference time. The vanilla SR3 and IDM, which employ 2,000 denoising steps, result in unacceptable inference times, leading to the common practice of utilizing accelerated sampling techniques from DDIM \cite{song2020denoising} to reduce computational costs.

While EDiffSR has considerably shortened its inference time compared to the original SR3, it still struggles to deliver SR results in real-time for practical applications. In contrast, our proposed E$^2$DiffSR significantly reduces inference time compared to previous diffusion-based methods, achieving a remarkable inference time of approximately 0.2 seconds for a $\times$4 SR task at a resolution of $512\times 512$. This represents a 15-fold improvement in inference efficiency over EDiffSR.

Regarding model parameters, our E$^2$DiffSR's parameter count, totaling 31.25 M, arises from the denoising network, conditioning network, and the decoder DEPSR in the first stage. Notably, this parameter count is only slightly larger than that of EDiffSR, while being significantly less than that of other diffusion-based methods. Overall, our proposal achieves an impressive balance between performance and computational efficiency, making it suitable for real-time SR tasks, while containing only a small number of parameters to save memory.

\section{Discussion}

In this section, we conduct ablation experiments to systematically investigate the contributions and impacts of various components of our proposed model. Additionally, we explore the configurations of the loss function.

\subsection{Analysis of Model Architecture}

In the encoder-decoder architecture, we innovatively proposed a differential prior encoding strategy, which is integrated with a conventional SR branch to alleviate the challenges associated with latent space pretraining. To validate the effectiveness of the differential prior encoding, we remove all connections between the DPDecoder and the SR branch, and the model containing only SR branch is denoted as \mbox{Model-I}. First, we evaluated the reconstruction performance of the SR branch without retraining its parameters. This enables evaluation of whether the encoded prior information is used to supplement high-frequency information for the SR task. Then, we utilized the same training setup to independently train the SR branch, which we designated as Model-I-Retrained.

Furthermore, we modified the decoder by removing the SR branch, resulting in a second variant, Model-II. This configuration degrade the autoencoder to one that is consistent with the conventional LDM. To accommodate various scale factors for the SR tasks within the decoder, we introduced a scale-aware feature modulation mechanism. To validate the effectiveness of this mechanism, we eliminated all modulation layers and directly concatenated all intermediate feature outputs from the DPDecoder with the features from the SR branch along the channel dimension, resulting in Model-III.

The results in Table \ref{tab:ablation-1} show that removing any component from the encoder-decoder architecture leads to a decline in performance. From the results of Model-I, it is evident that eliminating the differential prior information significantly degrades the reconstruction performance. While retraining the SR branch in Model-I-Retrained improves results, there remains a considerable gap compared to our proposed model. This highlights the critical role of differential prior information in achieving effective reconstruction. The substantial performance drop in Model-II demonstrates the difficulty of training traditional LDMs for the continuous-scale SR task. The results of Model-III show that our scale-aware feature modulation enhances the model’s multitasking capability. This mechanism allows flexible feature combination from both the SR branch and the DPDecoder under different scale factors, improving the model’s generalization performance across various tasks.

\begin{table}[t]
    \centering
    \renewcommand\arraystretch{1.2}
    \caption{Architecture Analysis of our Proposal with Different Components. Results are Tested on the AID-tiny Dataset with a Scale Factor of $\times$4. The Best Performance is Shown in \textbf{Bold}.}
    \setlength{\tabcolsep}{1pt}{
        \begin{tabular}{lcccccc}
            \Xhline{4\arrayrulewidth}
            Models & \makecell{Encoder-\\decoder} & \makecell{SR                                     \\branch}    & \makecell{Scale \\ modulation} & PSNR(dB) & FID & LPIPS \bigstrut \\
            \hline
            Model-I   & $\times$            & $\checkmark$ & $\times$           & 20.779 & 253.535 & 0.683 \bigstrut[t]\\
            Model-I-Retrained   & $\times$            & $\checkmark$ & $\times$  & 27.778 & 77.137 & 0.187 \\
            Model-II  & $\checkmark$ & $\times$            & $\checkmark$ & 23.826 & 147.701 & 0.290 \\
            Model-III & $\checkmark$ & $\checkmark$ & $\times$           & 28.100 & 43.479 & 0.098 \\
            Ours      & $\checkmark$ & $\checkmark$ & $\checkmark$ & \textbf{28.177} & \textbf{39.496} & \textbf{0.088} \bigstrut[b]\\
            \Xhline{4\arrayrulewidth}
        \end{tabular}%
    }
    \label{tab:ablation-1}%
\end{table}%

\subsection{Comparison of Regularization Methods}

We examined two commonly used regularization techniques to constrain the latent space: KL divergence regularization and VQ regularization. We compared the performance of these two methods, with the results presented in Table \ref{tab:ablation-2}. The first stage results reflect the reconstruction performance of the autoencoder, while the second stage results represent the final performance of the SR model. As shown in Table \ref{tab:ablation-2}, using KL divergence regularization leads to better results in both stages, demonstrating its superiority in optimizing the latent space and improving the overall performance of the model compared to VQ regularization. Therefore, we chose KL divergence regularization for our E$^2$DiffSR.

\begin{table}[t]
    \centering
    \renewcommand\arraystretch{1.2}
    \caption{Comparison of Regularization Methods. The Best Performance Is shown In \textbf{Bold}.}
    \setlength{\tabcolsep}{4pt}{
        \begin{tabular}{l|ccc|ccc}
            \Xhline{4\arrayrulewidth}
            \multirow{2}{*}{\makecell{Reg. \\methods}} & \multicolumn{3}{c|}{First Stage} & \multicolumn{3}{c}{Second Stage} \bigstrut\\
            \cline{2-7}          & PSNR(dB)  & FID   & LPIPS & PSNR(dB)  & FID   & LPIPS \bigstrut\\
            \hline
            KL-reg. & \textbf{28.177} & \textbf{39.496} & \textbf{0.088} & \textbf{28.148} & 61.321 & \textbf{0.175} \bigstrut[t]\\
            VQ-reg. & 27.094 & 46.603 & 0.126 & 27.213 & \textbf{56.061} & 0.196 \bigstrut[b]\\
            \Xhline{4\arrayrulewidth}
        \end{tabular}%
    }
    \label{tab:ablation-2}%
\end{table}%

\subsection{Analysis of Loss Function}

In the diffusion model training stage, the loss function consists of two terms, as shown in Eq.\,\eqref{eq:GaHG}. The first term $\|\hat{\mathbf{z}}_0 - \mathbf{z}_0\|_1$, referred to as the diffusion loss, ensures that the reverse process of the diffusion model generates samples consistent with the latent space. The second term $\|\mathopr{CondNet}(I_{\mathrm{LR}}^{\uparrow}) - \mathbf{z}_0\|_1$ in Eq.\,\eqref{eq:GaHG} is the knowledge distillation (KD) loss, which ensures that the conditional network generates conditions that match the true latent encoding, guiding the diffusion model during the denoising process. More importantly, we adopted a training strategy of using a limited number of reverse process time steps, where the denoising model traverses all reverse time steps instead of randomly sampling a time step, as is done in DDPM. To validate the effectiveness of this strategy, we compared it with the random time step sampling method used in DDPM. Table \ref{tab:ablation-3} presents the performance comparison of diffusion models trained with different loss configurations. The results clearly show that our chosen learning strategy and loss function settings significantly enhance the model’s performance.

\begin{table}[t]
    \centering
    \renewcommand\arraystretch{1.2}
    \caption{Discussion on Loss Functions of the Diffusion Model. The Best Performance Is shown In \textbf{Bold}.}
    \setlength{\tabcolsep}{8pt}{
        \begin{tabular}{lccc}
            \Xhline{4\arrayrulewidth}
            Models & PSNR(dB) & FID   & LPIPS \bigstrut\\
            \hline
            Random time step sampling & 23.660 & 106.956 & 0.338 \bigstrut[t]\\
            w/o KD loss & 28.037 & 68.692 & 0.195 \\
            w/o diffusion loss & 28.053 & 73.988 & 0.196 \\
            Ours  & \textbf{28.148} & \textbf{61.321} & \textbf{0.175} \bigstrut[b]\\
            \Xhline{4\arrayrulewidth}
        \end{tabular}%
    }
    \label{tab:ablation-3}%
\end{table}%

\section{Conclusions}
\label{sec:conclusion}

In this paper, we propose E$^2$DiffSR, a novel LDM-based framework for continuous-scale SR of RSIs. By encoding differential prior information between HR and LR images and utilizing a two-stage training approach, E$^2$DiffSR enhances the efficiency and quality of SR tasks. The integration of diffusion model learning in the latent space significantly improves the recovery of high-frequency details while maintaining computational efficiency. Extensive experiments on multiple remote sensing datasets demonstrate that E$^2$DiffSR surpasses existing SOTA regression-based, GAN-based, and diffusion-based SR approaches in both objective metrics and visual quality.

However, our approach has certain limitations. Although we have significantly reduced the number of sampling steps in the reverse diffusion process, there is still room for improvement in inference time compared to regression-based methods. Besides, we have not accounted for other complex degradation factors beyond downsampling. Therefore, in future work, we plan to introduce the consistency models that reduces the sampling of the reverse diffusion process to just one step to meet the demands of real-time applications. Furthermore, by considering more complex degradation factors, we aim to extend the proposed model to blind SR tasks, enhancing its applicability in real-world scenarios.

\balance


%

\appendices



\ifCLASSOPTIONcaptionsoff
    \newpage
\fi



\bibliographystyle{IEEEtran}
\bibliography{IEEEabrv,refs}
%



\balance

\end{document}